\documentclass[12pt]{article}
\usepackage{amsmath,amssymb,graphicx,axodraw,epsfig}
\newcommand{\T}{\mathcal{ T }}

\def \gsim{\mathrel{\vcenter
     {\hbox{$>$}\nointerlineskip\hbox{$\sim$}}}}

\newcommand{\thetabar}{\bar{\theta}}
\newcommand{\sigmabar}{\bar{\sigma}}
\newcommand{\alphadot}{{\dot{\alpha}}}
\newcommand{\betadot}{{\dot{\beta}}}

\def\bea{\begin{eqnarray}}
\def\eea{\end{eqnarray}}
\def\be{\begin{equation}}
\def\ee{\end{equation}}
\def\ba{\begin{array}}
\def\ea{\end{array}}
\def\nn{\nonumber}
\def\Z{{\bf Z}}
\def\a{&\hspace{-6pt}}
\def\csch{\,\mbox{csch}\,}

\font\tenrsfs=rsfs10
\font\sevenrsfs=rsfs7
\font\fiversfs=rsfs5
\newfam\rsfsfam
\textfont\rsfsfam=\tenrsfs
\scriptfont\rsfsfam=\sevenrsfs
\scriptscriptfont\rsfsfam=\fiversfs
\def\mathscr#1{{\fam\rsfsfam\relax#1}}

\begin{document}

\thispagestyle{empty}

\begin{center}
\hfill CERN-PH-TH/2004-223
\hfill NEIP/04-07
\hfill IFUP-TH/2004-33
\hfill HD-THEP-04-51

\begin{center}

\vspace{0.7cm}

{\LARGE \bf Gravitational quantum corrections in \\[2mm]
warped supersymmetric brane worlds}

\end{center}

\vspace{0.7cm}

{\large \bf T.~Gregoire$^{a}$, R.~Rattazzi$^{a}$\footnote{On leave 
from INFN, Pisa, Italy.}, C.~A.~Scrucca$^{b,a}$\\[2mm] 
A.~Strumia$^{c}$, E.~Trincherini$^{d,e}$}\\

\vspace{7mm}

${}^a${\em Physics Department, Theory Division, CERN, CH-1211 Geneva 23, Switzerland}
\vspace{.2cm}

${}^b${\em Institut de Physique, Universit\'e de Neuch\^atel, CH-2000 Neuch\^atel, Switzerland}
\vspace{.2cm}

${}^c${\em Dipartimento di Fisica, Universit\`a di Pisa, I-56126 Pisa, Italy}
\vspace{.2cm}

${}^d${\em Dipartimento di Fisica, Universit\`a di Milano Bicocca, I-20126 Milano, Italy}
\vspace{.2cm}

${}^e${\em Institut f{\"u}r Theor. Physik, Universit{\"a}t Heidelberg, D-69120 Heidelberg, Germany}
\vspace{.2cm}

\end{center}

\vspace{0.4cm}
\centerline{\bf Abstract}
\vspace{-0.1cm}
\begin{quote}
We study gravitational quantum corrections in supersymmetric theories with warped extra dimensions. 
We develop for this a superfield formalism for linearized gauged supergravity. We show that the 1-loop
effective K\"ahler potential is a simple functional of the KK spectrum  in the presence 
of generic localized kinetic terms at the two branes. We also present a simple understanding of our results 
by showing that the leading matter effects are equivalent to suitable displacements of the branes. 
We then apply this general result to compute the gravity-mediated universal soft mass $m_0^2$  in models where the 
visible and the hidden sectors are sequestered at the two branes. We find that the contributions 
coming from radion mediation and brane-to-brane mediation are both negative  in the minimal set-up, 
but the former can become positive if the gravitational kinetic term localized at the hidden brane has a sizable coefficient.
We then compare the features of the two extreme cases of flat and very warped geometry, and give an  outlook on the building 
of viable models.
\vspace{5pt}
\end{quote}

\newpage

\renewcommand{\theequation}{\thesection.\arabic{equation}}

\section{Introduction}\setcounter{equation}{0}

Low energy supersymmetry is arguably the best motivated extension of the Standard Model (SM). 
It solves the gauge hierarchy problem, has a natural dark matter candidate, and in the minimal 
scenario predicts gauge coupling unification. However, supersymmetry needs to be broken in order 
to give  weak scale masses to the superpartners.  In order for all the superpartners to be heavier 
than the SM particles, supersymmetry is typically broken in a hidden sector and transmitted to the 
SM by gravitational \cite{sugramediation} or gauge interactions 
\cite{Dine:1994vc} (for a review see \cite{Giudice:1998bp}). At low energy, the breaking of supersymmetry is encoded in soft supersymmetry 
breaking terms. A crucial point in designing a supersymmetry breaking scenario is to ensure 
that the soft scalar masses do not generate phenomenologically unacceptable Flavor Changing Neutral Currents 
(FCNC). The safest way of doing this is to generate soft masses in the IR, where the only flavor 
spurions are the Yukawa matrices. This insures that there will be a super-GIM mechanism  
suppressing FCNC. Gauge mediation is of this type, while ordinary gravity mediation is not, because 
the soft masses 
are affected by divergent gravity loops dominated in the UV. Another supersymmetry breaking transmission 
mechanism that is safe with respect to flavor is anomaly mediation \cite{Randall:1998uk,Giudice:1998xp}. 
In this scenario, supersymmetry breaking is transmitted via the super-Weyl anomaly, so that it is also dominated 
in the IR. However, the anomaly mediated contribution is parametrically smaller than the 
ordinary gravity mediated one. 
A way to suppress gravity mediation is to invoke an extra dimension of space. This allows to 
spatially separate the visible sector from the hidden sector where supersymmetry is broken. By locality, 
contact interactions between the two sectors are absent at tree level and are only generated 
by calculable gravity loops. The contribution to the soft parameters from these effective 
operators and the one from anomaly mediated have different radius dependence.  
For large enough  radius, anomaly mediation dominates,
leading to a sharp prediction for the soft terms. Unfortunately this sharp prediction entails tachyonic sleptons.
Thus pure anomaly mediation is not viable
and extra contributions should be invoked. A natural and interesting next-to-minimal scenario is obtained 
when the radius is small enough for  the finite gravitational loops
to compete with anomaly mediation effects \cite{Chacko:1999am}. In principle this can cure the tachyonic 
sleptons while keeping
an energy gap between the scale of mediation, the inverse radius $1/R$, and the five-dimensional (5D) 
quantum gravity scale $M_5$ at which presumably extra flavor breaking effects come into play. 
The presence of this energy gap allows to control the size of flavor violation in soft terms.

In the simplest situation of a flat geometry, the effective low-energy theory is described in terms of the 
visible and hidden sector chiral superfields $\Phi_i$ with $i=0,1$, the radion field $T$, whose scalar component 
vacuum expectation value (VEV) determines the radius $\langle T\rangle = \pi R$, and the 5D Planck scale $M_5$.
The tree-level kinetic function has the form\footnote{The K\"ahler potential is  given by  $K=- 3 \ln (-\Omega/3)$, in units of the 4D Planck mass. Throughout this paper, with an abuse of  nomenclature, we will often refer to $\Omega$ as the K\"ahler potential.}:
\begin{equation}
\Omega_{\rm tree} = - 3 M_5^3 \left(T + T^\dagger\right) +
\Phi_0^\dagger \Phi_0 + \Phi_1^\dagger \Phi_1 \;.
\end{equation}
The radius dependence of the non-local operators induced by gravity loops is completely fixed by simple 
power counting (notice that each graviton line brings in a factor of $1/M_5^3$).
The operators that are relevant for soft scalar masses are easily seen to have the form
\begin{equation}
\label{eq:flatkhalercorrections}
\Omega_{\rm 1-loop} \supset a \frac{\Phi_0^\dagger \Phi_0}{M_5^3 \left(T + T^\dagger\right)^3}
+ a \frac{\Phi_1^\dagger \Phi_1}{M_5^3 \left(T + T^\dagger\right)^3}
+ b \frac{\Phi_0^\dagger \Phi_0 \Phi_1^\dagger \Phi_1}{M_5^6 \left(T + T^\dagger \right)^4} \;.
\end{equation}
When supersymmetry is broken, the radion and the hidden sector chiral multiplets get in general non-zero $F$-terms, 
and the visible sector scalar fields get soft masses respectively from one of the first two terms and from the 
third term in ref.~\eqref{eq:flatkhalercorrections}.  The former give the so-called radion mediated contribution, 
and its coefficient $a$ was calculated for the first time in ref.~\cite{Gherghetta:2001sa}. The latter yields the 
brane-to-brane mediated contribution and its coefficient $b$ was recently calculated in 
refs.~\cite{Rattazzi:2003rj,Buchbinder:2003qu}. Both contributions turn out to be negative in the simplest situation.
However in ref.~\cite{Rattazzi:2003rj} the soft masses 
were computed also in a more general case, where the supergravity fields have non-vanishing localized kinetic terms. 
It was shown, in particular, that a kinetic term with a large coefficient at the hidden brane
 changes the sign of the 
radion-mediated contribution but not its size, whereas it does not change the sign of the brane-to-brane-mediated 
contribution but suppresses its size. It was then shown that in this situation a viable model of supersymmetry 
breaking with competing effects from gravity and anomaly mediation can be achieved.

The main qualitative effects of a large kinetic term on the hidden brane is to shift the spectrum of the Kaluza--Klein (KK) modes and localize their wave functions away from the hidden brane. This is somewhat similar to what happens 
in a warped geometry like the RS1 set-up of ref.~\cite{Randall:1999ee}. It is then conceivable that a warping 
of the geometry could lead to an acceptable pattern of gravity-mediated soft terms, representing perhaps 
a more natural and appealing substitute for the localized kinetic terms invoked in ref.~\cite{Rattazzi:2003rj}. 
The aim of this paper is to generalized the analysis of refs.~\cite{Rattazzi:2003rj,Buchbinder:2003qu} to warped 
geometries by computing the full effective K\"ahler potential in a supersymmetric version of the RS1 set-up. In 
order to investigate the quantitative relation between the effects of warping and localized kinetic terms, we shall 
moreover allow for arbitrary localized kinetic terms at the two branes in the warped case as well.

In the regime where the warping is significant, the expected form of the corrections can be deduced from the AdS/CFT 
correspondence \cite{maldacena:1997re}. According to this correspondence, all the physics of a RS1 model is equivalent to 
that of a 4D conformal field theory in which the conformal symmetry is non-linearly linearized in the IR and explicitly broken in 
the UV, and in which 4D gravity is gauged \cite{maldacena2,Gubser:1999vj,verlinde:1999fy,Arkani-Hamed:2000ds,Rattazzi:2000hs,Perez-Victoria:2001pa}. 
The coordinates that are most suitable for studying the holographic interpretation of the 
RS model are the ones where the metric is written as:
\begin{equation}
ds^2 = \frac{L^2}{ z^2} \left(dx_\mu d x^\mu + dz^2 \right)
\label{confmetr}
\end{equation}
where $L$ is the AdS radius length and where  boundaries at $z=z_0$ and $z=z_1$ with $z_0\ll z_1$ are assumed.
From the CFT point of view, the $z$ coordinate corresponds to a renormalization scale. In this respect the boundaries
at $z_0$ and $z_1$ are named respectively the UV brane and the IR brane. The presence of the UV 
brane corresponds to cutting off the CFT in the UV at the energy scale $1/z_0$ and to
gauging 4D gravity. Indeed the graviton zero mode is localized at the UV brane, and the effective 4D Planck mass is
$M^2= (M_5^3L^3)/z_0^2$. Notice that when $z_0\to 0$ the Planck mass diverges and 4D gravity decouples.
By a change of $x,z$ coordinates we can however always work with $z_0=L$, in which case we get the familiar RS
parametrization $M^2=M_5^3 L$. 
The presence of the IR brane at the position $z=z_1$ corresponds instead to a spontaneous 
breaking of conformal invariance in the IR at the energy $1/z_1$. The radion field of the 5D theory basically 
corresponds to fluctuations in the position $z_1$ of the IR brane  and can be interpreted  as the Goldstone 
boson of spontaneously broken dilatation invariance. Matter fields on the UV brane are identified 
with elementary fields coupled to the CFT through gravity and higher-dimensional operators suppressed by powers
of the UV cut-off $1/z_0$. On the other hand, compatibly with the interpretation of $z$ as an RG scale,
matter fields living at $z_1$ are interpreted as bound states with compositeness scale of order $1/z_1$.
In the limit where the position $z_0$ of the Planck brane is sent to $0$, the theory becomes conformal, implying, 
among other things, that the couplings of the radion are dictated by conformal invariance. It then follows that for 
$z_0\to 0$ the K\"ahler potential, including quantum corrections, must have its tree-level form. Any correction with 
a different dependence would explicitly break conformal invariance. Moving the Planck brane in, however,
makes the 4D graviton dynamical and breaks conformal symmetry explicitly. Due to this, non-trivial corrections 
to the K\"ahler potential are induced by graviton loops attached to the CFT.  These loops are cut-off at the KK 
scale $1/z_1$, playing the role of the scale of compositeness. The induced effects are therefore suppressed by 
powers of $1/(z_1 M)^2$.

To write the low energy effective theory of the supersymmetric RS model it is convenient to parameterize
the radion with a superfield $\mu$ whose scalar component VEV is precisely the position of the IR brane: 
$\langle \mu\rangle = 1/z_1$. The effective kinetic function
can then be written as
\begin{equation}
\Omega_{\rm tree} = - 3 M_5^3L^3(\frac{1}{L^2} -  {\mu^\dagger \mu} \Big)
+ \Phi_0^\dagger \Phi_0 + \Phi_1^\dagger \Phi_1  {\mu^\dagger \mu}{L^2} 
\label{treetree}
\end{equation}
Determining the form of the leading terms in the 1-loop action is more difficult than in the flat case, since there 
are now two length scales $L$ and $z_1$ instead of just one ($T$). Simple power counting
must therefore be supplemented by additional considerations. The use of the holographic pictures provides a 
very direct power counting insight. Apart from ``trivial'' UV divergent effects of the same form as the tree-level action, the 
calculable effects must be related to the explicit breakdown of conformal invariance due to the propagating
4D graviton. These corrections can be represented diagrammatically as in fig.~1. Keeping in mind that the quantum 
corrections are saturated by he only physical IR scale, i.e. the compositeness scale $\mu$, it is easy to power-count 
these effects to obtain
\begin{equation}
\Omega_{\rm 1-loop} \supset a_0 \frac {(\mu^\dagger \mu)^2 \Phi_0^\dagger \Phi_0}{k^2 M^2}
+ a_1 \frac {(\mu^\dagger \mu)^2 \Phi_1^\dagger \Phi_1}{k^2 M^2}
+ b \frac {(\mu^\dagger \mu)^2 \Phi_0^\dagger \Phi_0 \Phi_1^\dagger \Phi_1}{k^2 M^4} \;.
\label{powercount}
\end{equation}
where $k=1/L$. For instance the first term is  obtained by the second diagram in fig.~1 as the product of 
the following factors
\begin{equation}
N^2 \times \Phi_0^\dagger \Phi_0 \times \left (\frac{1}{M^2}\right )^2\times  (\mu\mu^\dagger)^2 \;,
\end{equation}
where the first factor $N^2\equiv (M_5L)^3$
counts the numerical coefficient in front of the CFT action \cite{maldacena2,Gubser:1999vj,verlinde:1999fy,Arkani-Hamed:2000ds,Rattazzi:2000hs,Perez-Victoria:2001pa}, the second and third factors 
are obvious, while the fourth factor is just dictated by dimensional analysis once $\mu$ is recognized as the only 
IR scale. Eq.~(\ref{powercount}) represents the leading effect due to 4D graviton loops. However the 
UV brane introduces also an explicit UV cut-off equal to $1/z_0\equiv 1/L$, so that we expect further corrections to 
eq.~(\ref{powercount}) supressed by powers of $\mu\mu^\dagger L^2$. These terms are unimportant for $\mu L\ll 1$,
i.e for large warping. However when the warping gets small these higher powers become crucial
to reproduce the flat case result.
Finally, upon supersymmetry breaking, there will again be a radion-mediated and 
a brane-to-brane-mediated contribution to the soft masses, and it is therefore important to compute 
the coefficients $a_{0,1}$ and $b$, and in particular their signs. 

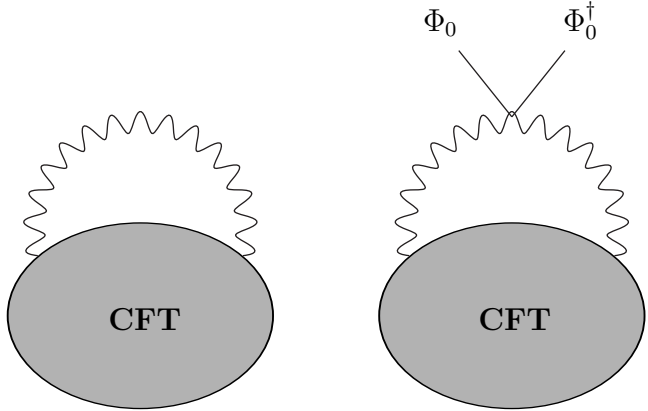
\begin{figure}
\begin{center}
\begin{picture}(210,160)(0,0)
\GOval(40,50)(35,50)(0){0.7}
\put(28,45){\bf{CFT}}
\GOval(180,50)(35,50)(0){0.7}
\put(168,45){\bf{CFT}}
\PhotonArc(40,83)(40,-15,195){4}{14.5}
\PhotonArc(180,83)(40,-15,195){4}{14.5}
\Line(180,125)(160,150)
\Line(180,125)(200,150)
\put(147,158){$\Phi_0$}
\put(200,158){$\Phi_0^\dagger$}
\end{picture}
\end{center}
\caption{\em The two kinds of diagrams that are responsible for the effective operators 
that we are interested in from the CFT point of view.}
\label{}
\end{figure}

The results of the explicit computation  we shall present in the next sections show that in the 
absence of localized kinetic terms, the coefficients $a_{0,1}$ and $b$ are positive and the corresponding 
soft masses squared are negative. This means that
a significant warping of the geometry, while qualitatively similar to, is quantitatively different than 
having large localized kinetic terms in flat space. In the presence of a single sizable localized kinetic 
term on one of the two branes, which is then identified with the hidden sector, the situation is found to 
change as follows. If the hidden sector is at the UV brane, the effect of the localized kinetic term amounts 
basically to a rescaling of the 4D Planck mass, and the coefficients $a_1$ and $b$ that are 
relevant for the radion-mediated and brane-to-brane-mediated effects get both suppressed without changing 
sign. If instead the hidden sector is at the IR brane, a non-vanishing localized kinetic term tends to change 
the sign of the coefficient $a_0$ relevant for radion mediation, and to suppress the size of the coefficient 
$b$ relevant for brane-to-brane mediation while preserving its sign. This last case is therefore potentially as 
viable as the flat case. However there is a limitation on the size of the coefficient of the localized kinetic 
term on the IR brane: if it is too large then the radion becomes a ghost \cite{Luty:2003vm}. In order to 
understand whether and to what extent a warping of the geometry can be a helpful and appealing supplement 
to localized kinetic terms, it is therefore necessary to perform a detailed analysis at finite warping and localized 
kinetic terms.

There are various techniques that can be used to perform the calculation of the gravitational loop effects 
discussed above in the general case of finite warping. One possible approach is the off-shell component 
formalism of Zucker \cite{Zucker:1999ej,Zucker:1999fn,Zucker:2000ks}, which was used in the flat case 
in ref.~\cite{Rattazzi:2003rj}. This approach is somewhat inconvenient in the warped case for two reasons. 
First, the formalism is plagued with singular products of distributions that are hard to deal with in 
warped space. Second, the trick used by ref.~\cite{Rattazzi:2003rj} of considering a background with 
a non-zero VEV for the $F$-term of the radion multiplet and calculating the induced potential instead of 
the correction to the K\"ahler potential, is also hard to generalize to the warped case. Another possible 
approach is the superfield formalism employed in ref.~\cite{Buchbinder:2003qu}, where only half of the bulk 
supersymmetry is manifest. This turns out to be easier to generalize to the warped case, and we shall 
therefore use it as the basic framework to set up the computation. In the end, however, it turns out that all 
the information that is needed to derive the result is effectively contained in the spectrum of a single
massless scalar  with arbitrary localized kinetic 
terms. In fact, the calculation of the full $1$-loop correction to the K\"ahler potential is quite analogous
to  that of the effective potential for this simple system.

The paper is organized as follows. In sec.~2 we describe more precisely the context of our computation, namely
the warped supersymmetric brane worlds based on the RS1 geometry. In sec.~3 we generalize the linearized superfield
formalism of ref.~\cite{Buchbinder:2003qu} to the warped case, apply it to set up the supergraph computation of 
the gravitational loop effects that we want to compute, and study the structure of the latter to show that it 
effectively maps to a computation within a simple free theory of a real scalar field with localized kinetic terms.
In sec.~4 we concretely perform the computation of the loop effects and discuss the results and their implications
on the gravity-mediated soft mass terms. In sec.~5 we present a simple argument allowing to relate the form
of the matter-dependent effects to the matter-independent one. In sec.~6 we discuss the application of our results 
to model building. Finally, app.~A contains some useful technical details concerning the linearized superfield description 
of warped models. The readers who are not interested in the details concerning a rigorous set up of the computation
within the superfield approach can skip sec.~\ref{sec:superfield} and app.~A and take sec.~4 as a starting point 
for the computation.

\section{Warped supersymmetric brane worlds}\setcounter{equation}{0}

In this section, we shall describe more precisely the context of our computation and introduce the basic notation
we will use throughout the paper. Our starting point is the supersymmetric version of the 5D RS1 geometry 
\cite{Zucker:2000ks,Gherghetta:2000qt,Falkowski:2000er,Bergshoeff:2000zn,Altendorfer:2000rr,Bagger:2002rw,Lalak:2003fu,Zucker:2003qv}. 
The theory is constructed by performing a gauging of 5D supergravity compactified on the orbifold $S^1/\Z_2$. 
Denoting by $y$ the internal coordinate and by $y_0 = 0$ and $y_1 = \pi R$ the two fixed-points, the Lagrangian 
takes the form:
\bea
{\cal L} = {\cal L}_5 + \delta_0(y) {\cal L}_0 + \delta_1(y) {\cal L}_1 \;,
\label{Lagrangian}
\eea
where $\delta_i(y) = \delta(y-y_i)$ and
\footnote{There are two formulations of the supergravity extension of the RS model (presented respectively in 
\cite{Gherghetta:2000qt,Falkowski:2000er,Bergshoeff:2000zn} and \cite{Zucker:2000ks,Altendorfer:2000rr}) 
that are equivalent but differ by a singular gauge transformation \cite{Bagger:2002rw,Lalak:2003fu,Zucker:2003qv}. 
This singular transformation makes the second formulation hard to deal with, and we therefore use the first.}
\bea
{\cal L}_5 \a=\a \sqrt{g_5} \Big\{\!\!-\! \Lambda_5\! 
- \frac 12 M_5^3 \Big[{\cal R}_5 + i \bar \Psi_{\!M} \Big(\Gamma^{MRN}\! D_R 
\!-\! \frac {3i}2 k \epsilon(y) \Gamma^{MN}\Big)\Psi_{\!N} 
+ \frac 12 F_{MN}^2 \Big] 
+ \cdots \! \Big\} \;,\nn \\
{\cal L}_i \a=\a \sqrt{g_4} \Big\{\!\!-\! \Lambda_i - \frac 12 M_i^2 \Big({\cal R}_4 
+ i \bar \Psi_\mu \gamma^{\mu\rho\nu} D_\rho \Psi_\nu \Big) 
+ \Big(|\partial_\mu \phi_i|^2 + i \bar \psi_i \gamma^\mu D_\mu \psi_i \Big) 
+ \cdots \! \Big\} \;. \nn
\eea
In this expression $M_5$ is the 5D fundamental scale, $R$ is the radius of the 
compact extra dimension, $k$ is a curvature scale and $M_{0,1}$ are two 
scales parametrizing possible localized kinetic terms for the bulk fields, 
whereas $\Lambda_5$ and $\Lambda_{0,1}$ are a bulk and two boundary cosmological 
constants that are tuned to the following values
\be
\Lambda_5 = - 6 M_5^3 k^2 \;,\;\;
\Lambda_0 = - \Lambda_1 = 6 M_5^3 k \;.
\ee
We assume as usual that the compactification scale $1/R$ and the curvature $k$ are 
much smaller than the fundamental scale $M_5$, so that we can reliably use the above 
5D supergravity theory as an effective description of the physics of the model and 
neglect higher-dimensional operators.

The above theory has a non-trivial supersymmetric warped solution defining a 
slice of an ${\rm AdS}_5$ space that is delimited by the two branes at $y=y_{0,1}$. 
Defining $\sigma(y) = k |y|$, the background is given by 
\be
g_{MN} = e^{-2 \sigma} \eta_{\mu \nu} \delta^\mu_M \delta^\nu_N 
+ \delta^y_M \delta^y_N \;,\;\; \Psi_M = 0 \;,\;\; A_M = 0 \;.
\label{background}
\ee
Notice that the 4D geometry is flat at every point $y$ of the internal dimension,
but the conformal scale of the metric varies exponentially with it. This gives rise 
to a dependence of physical effective energy scales on $y$.

\subsection{Effective theory}

At energies much below the compactification scale, the fluctuations around the
above background solution are described by a 4D  supergravity 
with vanishing cosmological constant. The fields of this effective theory are the 
massless zero mode of the 5D theory and fill out a supergravity multiplet 
$G = (h_{\mu \nu},\psi_\mu)$ and a radion chiral multiplet $T = (t, \psi_t)$. 
They are defined by parametrizing the fluctuations around the background as follows:
\bea
g_{\mu\nu} \a=\a \exp \Big(\!-\!2 \sigma \frac {{\rm Re}\,t}{\pi R}\Big) 
\big(\eta_{\mu \nu} + h_{\mu \nu} \big) \;,\;\;
g_{\mu y} = 0 \;,\;\; g_{yy} = \Big(\frac {{\rm Re}\,t}{\pi R}\Big)^2 \;; \nn \\
\Psi_\mu \a=\a \psi_\mu \;,\;\; \Psi_y = \frac {\sqrt{2}}{\pi}\psi_t \;;\;\;
A_\mu = 0 \;,\;\; A_y = \frac {\sqrt{3}}{\sqrt{2}\pi}\,{\rm Im}\,t\;.
\label{fluctuations}
\eea
Substituting eqs.~(\ref{fluctuations}) into eq.~(\ref{Lagrangian}) and 
integrating over the internal dimension, one finds the following effective K\"ahler potential \cite{Luty:2000ec}:
\bea
\Omega(T+T^\dagger,\Phi_i,\Phi_i^\dagger) \a=\a 
- 3 \frac {M_5^3}{k} \Big(1 - e^{- k (T + T^\dagger)} \Big) \nn \\ \a\;\a 
+\, \Omega_0(\Phi_0,\Phi_0^\dagger) 
+ \Omega_1(\Phi_1,\Phi_1^\dagger) \,e^{- k (T + T^\dagger)} \;.\qquad
\label{kahlereff}
\eea
The first term is the matter-independent contribution from the bulk, whereas 
the last two terms are the matter-dependent contributions from the branes
\be
\Omega_i(\Phi_i,\Phi_i^\dagger) = - 3 M_i^2 +\Phi_i^\dagger  \Phi_i  + \dots \;.
\label{kahlerboundary}
\ee
The dots denote irrelevant operators involving higher powers of 
$ \Phi_i^\dagger \Phi_i$. Note that there is a limit on the possible size of $M_1$, because if 
$M_1^2 > M_5^3/k$, the radion, parametrized by $\exp(-k T)$, becomes a ghost \cite{Luty:2003vm}. 
Superpotentials $W_i$ localized on the branes give rise to the following effective superpotential  at low energy:
\bea
W(T,\Phi_i) = W_0(\Phi_0) + W_1(\Phi_1)\, e^{- k T} \;.
\label{supereff}
\eea
Finally, the effective Planck scale can be read off from the part of (\ref{kahlereff})
 after substituting $T$  with its VEV $\pi R$. One finds, assuming vanishing matter VEVs,
\be
M^2 = \frac {M_5^3}{k} \Big(1 - e^{- 2 \pi k R} \Big) 
+ M_0^2 + M_1^2\, e^{- 2 \pi k R} \;.
\ee

Assume now that the field $\Phi_0$ represents collectively the fields of the visible sector,
while $T$ and $\Phi_1$ represent the hidden sector. It is evident that the above tree level action does not induce any soft terms even after supersymmetry is broken. Notice that the result would be the same in the reversed situation where $\Phi_1$
represents the visible sector. This is because the radion couples  to $\Phi_1$
as a conformal compensator: tree level effects can be easily seen to cancel through a
field redefinition  $\Phi_1'=\Phi_1 e^{-kT}$. Therefore the above tree Lagrangian realizes the sequestering of the hidden sector.  Soft terms can only be induced by calculable quantum corrections.

\subsection{Loop corrections}

The corrections to the K\"ahler potential of the effective theory that are induced 
by loops of bulk modes come in two different classes. The first class represents 
a trivial renormalization of the local operators corresponding to the 
classical expression (\ref{kahlereff}).  As we just explained, terms of this form do not mediated soft masses: although UV divergent (or better, uncalculable), this correction
is uninteresting.
 A second class corresponds instead to new  effects that 
have a field dependence different from the one implied by locality and general covariance in (\ref{kahlereff}). By definition these effects are genuinely non-local and, therefore, finite and calculable. These corrections can be 
parametrized in the following general form:
\be
\Delta \Omega(T+T^\dagger,\Phi_i,\Phi_i^\dagger) = 
\sum_{n_0,n_1=1}^\infty C_{n_0,n_1}(T + T^\dagger) ( \Phi_0^\dagger \Phi_0)^{n_0} 
( \Phi_1^\dagger \Phi_1)^{n_1} \;.
\label{corrkahlereff}
\ee
The functions $c_{n_0,n_1}$ control the leading effects allowing the transmission 
of supersymmetry breaking from one sector to the other, and can be computed along the 
same lines as for the flat case, which was studied in refs.~\cite{Rattazzi:2003rj} and 
\cite{Buchbinder:2003qu}. 

Unfortunately the trick that was used in ref.~\cite{Rattazzi:2003rj}, namely computing 
the effective potential at $F_T \neq 0$ and deducing 
from it the form of eq.~(\ref{corrkahlereff}), cannot be 
generalized in a straightforward way to the warped case. In the flat case, a consistent 
tree-level solution with $F_T\not = 0$ and flat 4D (and 5D) geometry could be found by 
simply turning on constant superpotentials $W_0$ and $W_1$ at the boundaries. Since  the interesting terms in the K\"ahler potential are associated to terms quadratic 
in $F_T$ in the effective potential, and since $F_T\propto W_0 + W_1$, it was enough to work with infinitesimal $W_{0,1}$ and 
calculate the effective potential at quadratic order in $W_{0,1}$. Notice, by the way, 
that boundary superpotentials are just the only local, zero-derivative deformation that 
is available in 5D Poincar\'e supergravity. The situation is drastically modified in 5D 
AdS supergravity. In this case, by turning on boundary superpotentials we have that 
\cite{Bagger:2003dy,Bagger:2003fy,Derendinger:2004kf}: 1) the radius is stabilized, 2) the 4D metric of the 4D slices 
(and of the low energy effective theory) becomes ${\rm AdS}_4$, 3) there is a (compact) 
degeneracy of vacua associated to the VEV of the graviphoton component $A_5$. At all points 
the scale of supersymmetry breaking is subdominant to the scale of ${\rm AdS}_4$ curvature 
and at a special point supersymmetry is restored. The last property can be quantified by 
the deviation $\delta m_{3/2}$ of the gravitino mass from its supersymmetric value, which is
equal to the curvature $1/L_4$ in ${\rm AdS}_4$. One obtains $\delta m_{3/2} L_4\leq \omega^2$, 
where $\omega$ is the ${\rm AdS}_5$ warp factor. In principle one could go ahead and calculate 
corrections to the effective potential in this background and read back from it the effective 
K\"ahler potential. However the 4D curvature, as we said, cannot be treated as a subleading 
effect and this complicates both the calculation and the indirect extraction of the K\"ahler 
potential.

Rather than trying to encompass this difficulty, we will generalize to the warped case the 
linearized superfield approach that was used in \cite{Buchbinder:2003qu} and in which the K\"ahler 
potential is calculated directly. This generalization is interesting on its own, and we will 
present it in detail in the next section. However, it turns out that the result that it produces 
for the correction to the effective K\"ahler potential is a very intuitive and obvious generalization 
of those derived in \cite{Rattazzi:2003rj} and \cite{Buchbinder:2003qu} for the flat case: one 
has just to replace the flat space propagators with the corresponding warped space ones. 
We shall prove this general result in a rigorous way with superfield techniques in the next section. The actual computation is
postponed to a subsequent section.

\section{Superspace description}\setcounter{equation}{0}
\label{sec:superfield}

A convenient way of performing loop calculations in supersymmetric theories is to use 
supergraph techniques. By calculating loop diagrams directly in terms of superfields, 
the number of graphs is greatly reduced, and various cancellations between graphs that 
are insured by supersymmetry are guaranteed to happen. Unfortunately, the use of this 
technique in theories with more than four space-time dimensions is not straightforward, 
because the amount of supersymmetry is higher than in four dimensions. From a 4D 
perspective, there are in this case several supercharges, and the simultaneous realization 
of the associated symmetries requires a superspace with a more complex structure. However, 
it is still possible to manifestly realize one of the 4D supersymmetries 
on a standard superspace, at the expense of losing manifest higher-dimensional Lorentz 
invariance \cite{Marcus:1983wb,Arkani-Hamed:2001tb,Arkani-Hamed:1999pv,Linch:2002wg}. 
In this way, only a minimal $N=1$ subgroup of the extended higher-dimensional 
supersymmetry will be manifest, but this turns out to be enough for our purposes.

In fact, writing higher-dimensional supersymmetric theories in term of $N=1$ superfields
not only simplifies loop computations, but makes it also easier to write down supersymmetric couplings 
between bulk and brane fields. As explained in ref.~\cite{Mirabelli:1998aj}, the way of doing 
this is to group the higher-dimensional supermultiplets into subsets that transform under 
an $N=1$ subgroup of the full higher-dimensional supersymmetry. Brane couplings can then
be written using the known 4D $N=1$ supersymmetric couplings. Splitting higher-dimensional 
multiplets in different $N=1$ superfields does just that, without having to look explicitly at the 
higher-dimensional supersymmetry algebra. Also, the couplings in component form often 
involve ambiguous products of $\delta$ functions that arise when auxiliary fields are 
integrated out. Using superfields this problem is avoided, because auxiliary fields are 
never integrated out. The drawback of the superfield approach is that, since higher-dimensional 
Lorentz invariance and the full supersymmetry are not manifest, one has to more or less guess the 
Lagrangian, and check that it reproduces the correct higher-dimensional Lagrangian in components.

This technique has been used for studying linearized 5D supergravity in flat space in 
ref.~\cite{Linch:2002wg}, and the resulting formalism has been successfully applied 
in ref.~ \cite{Buchbinder:2003qu} to compute gravitational quantum corrections in 
orbifold models, with results that agree with those derived in ref.~\cite{Rattazzi:2003rj}
by studying a particular component of the corresponding superspace effective operators.
In the following, we will briefly review the approach of refs.~\cite{Linch:2002wg,Buchbinder:2003qu} 
for the case of a flat extra dimension, and then generalize it to the case of a warped extra 
dimension.

\subsection{Bulk lagrangian for a flat space}

The propagating fields of 5D supergravity consist of the graviton $h_{MN}$, the graviphoton 
$B_M$ and the gravitino $\Psi_M$, which can be decomposed into two two-components Weyl spinors 
$\psi^+_M$ and $\psi^-_M$. These fields can be embedded into a real superfield $V_m$, a complex 
general superfield $\Psi_\alpha$, and two chiral superfields $\mathcal{T}$ and $\Sigma$, according
to the following schematic structure:
\begin{eqnarray}
V_m \a=\a \theta \sigma_n \thetabar (h_{mn} - \eta_{mn} h) 
+ \bar{\theta}^2 \theta \psi^+_m + \cdots \;, \\
\Psi_\alpha \a=\a \thetabar \sigma^m \left(B_m + i h_{m y}\right) 
+ \theta \sigma_m \thetabar \psi^-_m + \thetabar^2 \psi_y^+ + \cdots \;, \\
\mathcal{T} \a=\a h_{yy} + i B_y + \theta \psi_y^- + \cdots \;, \\
\Sigma \a=\a s + \cdots \;.
\end{eqnarray}
The dots denote higher-order terms involving additional fields, which are either 
genuine auxiliary fields or fields that are a priori not but eventually turn out to be 
non-propagating \cite{Linch:2002wg}. We also need to introduce a real 
superfield $P_\Sigma$ acting as a prepotential for the chiral conformal compensator 
$\Sigma$: $\Sigma = -1/4 \bar{D}^2 P_\Sigma$. This introduces yet more non-propagating fields.

Using the above fields, it is possible to construct in an unambiguous way a linearized 
theory that is invariant under infinitesimal transformations of all the local symmetries 
characterizing a 5D supergravity theory on an interval. These linearized gauge transformations 
consist of the usual 4D superdiffeomorphisms, which are parametrized by a general complex 
superfield $L_\alpha$, and the additional transformations completing these to 5D 
superdiffeomorphims, which are parametrized by a chiral multiplet $\Omega$. The corresponding 
linearized gauge transformations of the superfields introduced above are given by 
\bea
\label{eq:transfflat}
\delta V_m \a=\a -\frac{1}{2} \sigmabar_m^{\alpha \alphadot}
\left(\bar{D}_\alphadot L_\alpha - D_\alpha \bar{L}_\alphadot \right) \nn \\[-1.5mm]
\delta \Psi_\alpha \a=\a \partial_y L_\alpha - \frac{1}{4} D_\alpha \Omega \nn \\[0.5mm]
\delta P_\Sigma \a=\a D^\alpha L_\alpha + \text{h.c} \nn \\[1mm]
\delta \mathcal{T} \a=\a \partial_y \Omega \;.
\eea
As usual, $L_\alpha$ also contains conformal transformations that extend the 4D
super-Poincar\'e group to the full 4D superconformal group, but these extra 
symmetries are fixed by gauging away the compensator multiplet $\Sigma$.

The Lagrangian for linearized 5D supergravity in flat space can be constructed 
by writing the most general Lagrangian that is invariant under the above linearized
gauge transformations. This fixes the Lagrangian up to one unknown constant that can 
be determined by imposing that the component form be invariant under 5D Lorentz 
transformations. The result is
\begin{eqnarray}
\label{eq:linflatsugra}
\mathcal{L} = M_5^3 \! \int \! d^4 \theta \a\!\bigg\{\!\a
\frac 12 V^m K_{mn} V^n -\frac{1}{3} \Sigma^\dagger \Sigma + \frac{2 i}{3} 
\left(\Sigma - \Sigma^\dagger\right) \partial^m V_m \nn \\
\a\!\!\a - \frac{1}{2}\Big[\partial_y V_{\alpha \alphadot} 
- \big(\bar{D}_\alphadot \Psi_\alpha - D_\alpha \bar{\Psi}_\alphadot \big) \Big]^2
+ \frac{1}{4} \Big[\partial_y P_\Sigma 
- \big(D^\alpha \Psi_\alpha + \bar{D}_\alphadot \bar{\Psi}^\alphadot \big) \Big]^2 \nn \\
\a\!\!\a -\frac{1}{2}\Big[\mathcal{T}^\dagger \big(\Sigma + 2 i \partial_m V^m\big) 
+ \text{h.c.} \Big] \bigg\} \;,
\eea
where
\begin{equation}
\label{eq:kmn}
K_{mn} = \frac{1}{4} \eta_{nm} D^\alpha \bar{D}^2 D_\alpha 
+ \frac{1}{24} \sigmabar_m^{\alphadot \alpha} \sigmabar_n^{\betadot \beta} 
\left[D_\alpha,\bar{D}_\alphadot \right] \left[D_\beta,\bar{D}_\betadot \right] 
+ 2 \partial_m \partial_n \;.
\end{equation}
The first line of \eqref{eq:linflatsugra}  has the same form as the usual 4D linearized supergravity Lagrangian. 
To obtain the component Lagrangian, one chooses a suitable Wess--Zumino type of gauge, and 
eliminates all the auxiliary fields. By doing so, one correctly reproduces the 
linearized Lagrangian of 5D supergravity, with in addition some extra fields that do not 
propagate but have the dimensionality of propagating fields \cite{Linch:2002wg}.  The above construction can be generalized 
to the $S_1/{\Z}_2$ orbifold in a straightforward way, by assigning a definite ${\Z}_2$ parity 
to each multiplet: $V_m$, $\Sigma$ and $\mathcal{T}$ are even, whereas $\Psi_\alpha$ is odd. 

The 5D Lagrangian \eqref{eq:linflatsugra} can be written in a physically more transparent 
form by using a complete set of projectors that defines the different orthogonal components of the real superfield $V_m$ 
with {\it superspin} $0$, $1/2$, $1$ and $3/2$. The projectors are given by \cite{Gates:2003cz,Gates:1983nr,Gregoire:2004ic}: 
\begin{eqnarray}
\Pi^{mn}_0 \a=\a \Pi_L^{mn} P_C \;, \\
\Pi_{1/2}^{mn} \a=\a \frac{1}{48} \frac{1}{\square} 
\sigma^m_{\alpha \alphadot} \sigma^n_{\beta \betadot} 
\left[D_\alpha,\bar{D}_\alphadot \right]\left[D_\beta,\bar{D}_\betadot \right] 
+ \Pi_L^{mn} P_T + \frac{1}{3} \Pi_0^{mn} \;, \\[0.5mm]
\Pi_1^{mn} \a=\a \Pi_T^{mn} P_C \;,\\
\Pi_{3/2}^{mn} \a=\a  -\frac{1}{48} \frac{1}{\square}
\sigma^m_{\alpha \alphadot} \sigma^n_{\beta \betadot} 
\left[D_\alpha,\bar{D}_\alphadot \right]\left[D_\beta,\bar{D}_\betadot \right] 
+ \eta^{mn} P_T - \Pi_L^{mn} + \frac{2}{3} \Pi_0^{mn} \;,
\end{eqnarray}
in terms of the transverse and chiral projectors on vector superfields, 
which are given by 
\begin{equation}
P_T = -\frac{1}{8} \frac{D^\alpha \bar{D}^2 D_\alpha}{\square} \;,\;\;
P_C = \frac{1}{16} \frac{D^2 \bar{D}^2 +\bar{D}^2 D^2}{\square} \;,
\end{equation}
and the transverse and longitudinal projectors acting on vector indices, given by
\begin{equation}
\Pi_T^{mn} = \eta_{mn} - \frac{\partial_m \partial_n}{\square} \;,\;\;
\Pi_L^{mn} = \frac{\partial_m \partial_n}{\square} \;.
\end{equation}
The kinetic operator \eqref{eq:kmn} can then be written as 
\be
K^{mn} = - 2\, \square \Big(\Pi_{3/2}^{mn} - \frac 23\,\Pi_{0}^{mn}\Big)\;,
\ee 
Using the above complete set of superspin projectors, we can split the 
field $V_m$ into four orthogonal parts $V_0$, $V_{1/2}$, $V_1$ and $V_{3/2}$.
The first three transform non-trivially under local super-diffeomorphism, 
but not the last one, which is invariant. The Lagrangian \eqref{eq:linflatsugra}
can then be equivalently rewritten as
\bea
\label{eq:linflatsugraproj}
\mathcal{L} = M_5^3 \! \int \! d^4 \theta \a\!\bigg\{\!\a 
\!- V_{3/2}^m \big(\square + \partial_y^2\big) V_{3/2 m} 
- \frac 23 \Big[\partial_m V_0^m - \frac i2 \big(\Sigma - \Sigma^\dagger\big) \Big]^2 \nn \\
\a\!\!\a\!\! - \frac 12 \Big[\partial_y \big(V_{0}^{\alphadot \alpha} 
+ V_{1/2}^{\alphadot \alpha} + V_{1}^{\alphadot \alpha} \big)
- \big(\bar{D}_\alphadot \Psi_\alpha - D_\alpha \bar{\Psi}_\alphadot \big)\Big]^2 \\
\a\!\!\a\!\! + \frac{1}{4} \Big[\partial_y P_\Sigma - \big(D^\alpha \Psi_\alpha 
+ \bar{D}_\alphadot \bar{\Psi}^\alphadot \big) \Big]^2 \!\!
+ i \big(\mathcal{T} \!-\! \mathcal{T}^\dagger\big) \Big[\partial_m V_0^m 
\!-\! \frac i2 \big(\Sigma - \Sigma^\dagger\big) \Big] \bigg\} \;.\nn
\eea
We can see very clearly in this language why the compensator is needed. The kinetic Lagrangian 
for the gauge-invariant component $V_{3/2}^m$ is non local, due to the singular form of 
$\Pi_{3/2}^{mn}$. This non-local part is cancelled by a similar non-local part coming 
from the kinetic Lagrangian of the gauge-variant component $V_{0}^m$. The non-trivial variation 
under gauge transformations of this term is then compensated by that of $\Sigma$. Therefore the kinetic 
term of linearized  4D supergravity,  the first line of eq.~(\ref{eq:linflatsugraproj}), decomposes 
as the sum of two invariant terms respectively of maximal (3/2) and minimal (0) superspin. 
This is fully analogous to the situation in ordinary Einstein gravity, where the linearized kinetic 
term decomposes as the sum of spin 2 and spin 0 components. Notice also that, like in Einstein gravity, 
$V_{3/2}^m$ being the only component of maximal superspin, it cannot mix to any other component.  

We can verify that the correct 4D $N=1$ effective Lagrangian is obtained for the zero modes, 
by taking the even fields to depend only on the four dimensional coordinates and integrating 
over the extra dimension with radius $R$. To be precise, we use a hat to distinguish the 4D 
zero mode of each field from the corresponding 5D field itself. The result is:
\bea
\label{eq:zeromodeflat}
\mathcal{L}_{\rm eff} = 2 \pi R M_5^3 \! \int \! d^4 \theta \a\!\bigg\{\!\a
\!- \hat V_{3/2}^m \big(\square \big) \hat V_{3/2 m} 
- \frac 23 \Big[\partial_m \hat V_0^m 
- \frac i2 \big(\hat \Sigma - \hat \Sigma^\dagger\big) \Big]^2 \nn \\
\a\!\!\a \!\! +\, i \big(\hat{\mathcal{T}} - \hat{\mathcal{T}}^\dagger\big) 
\Big[\partial_m \hat V_0^m - \frac i2 \big(\hat \Sigma - \hat \Sigma^\dagger\big) 
\Big] \bigg\} \;.
\eea
It can be verified (trivially for the $\hat V_m$-independent terms) that this is indeed 
the quadratic expansion of the 4D supergravity Lagrangian, which can be written in 
terms of the 4D conformal compensator $\phi = \exp\,(\hat \Sigma/3)$ and the full
4D radion field $T = \pi R (1 + \hat{\mathcal{T}})$ as
\begin{equation}
\mathcal{L}_{\rm eff} = -3 M_5^3 \int d^4 \theta \,\big(T + T^\dagger\big)
\phi^\dagger \phi \;.
\end{equation}
In this expression, the $d^4 \theta$ integration is in fact an abbreviated notation 
for taking the $D$-term in a covariant manner. In particular, factors of the metric 
should be included. This result agrees with what was found in \cite{Luty:1999cz}.

\subsection{Bulk Lagrangian for warped space}

We now turn our attention to the case of warped space. $\text{AdS}_5$ is not a solution 
of the ordinary, ungauged supergravity Lagrangian, which does not admit a cosmological 
constant. To have a cosmological constant term, a $U(1)_R$ subgroup of the $SU(2)_R$ 
symmetry must be gauged by the graviphoton. However, because we will restrict ourselves 
to the quadratic Lagrangian of supergravity, the gauging of the $U(1)_R$ will not be apparent 
in our formalism. Within this gauged theory, we then assume a fixed background defined by 
eq.~\eqref{background} and look for the quadratic Lagrangian for the fluctuations around that background. 

The first important thing that we want to show is that the fluctuation Lagrangian can be 
written in terms of ordinary 4D superfields. To do so we start by considering the quadratic 
Lagrangian for 5D supergravity. At the local level there are two supersymmetries. However 
the boundary conditions on $S_1/\Z_2$ are such that globally there remains at most one 
supersymmetry, regardless of there being 5D curvature. As it has been discussed in several 
papers, the locally supersymmetric RS1 model preserves one global supersymmetry $Q_\alpha^G$. 
Technically this means that there exists one killing spinor $\xi$ over the RS1 background. 
In bispinor notation, and introducing a generic constant Weyl spinor $\eta^\alpha$,
this Killing spinor has the form \cite{Bergshoeff:2000zn}
\be 
\xi =\left (\begin{array}{c}e^{-\sigma/2} \eta^\alpha \\ 0 \end{array}\right) \;.
\ee
The generator $Q_\alpha^G$ of the corresponding global supersymmetry is defined in terms of 
the generators of the local 5D supersymmetries, denoted in bispinor notation by 
$Q^L\equiv ({Q_2^L}^\alpha, {{\bar Q}^L}_{1\dot\beta})$, by the equation
$\eta^\alpha Q_\alpha^G = \xi {\bar Q}^L = e^{-\sigma/2} \eta^\alpha {Q^L}_{1\alpha}$,
which implies
\be
Q_\alpha^G = e^{- \sigma/2} Q_{1\alpha}^L \;.
\label{global}
\ee
We can now realize $Q^G$ and the rest of the global 4D super-Poincar\'e group ($P_\mu=i\partial_\mu$ 
plus Lorentz boosts) over ordinary flat 4D superspace:
\bea
Q_{\alpha}^G \a=\a \frac{\partial}{\partial \theta^\alpha} 
- i \sigma^a_{\alpha \alphadot} \bar \theta^{\dot \alpha} {\delta_a}^\mu \partial_\mu \;,
\label{Q} 
\eea
In this realization of our field space, the fifth coordinate $y$ is just a label upon which our 4D superfields 
$S(x,y,\theta)$ depend. Of course 5D covariance is never manifest in this formulation of the theory and the 
correct action is obtained via the explicit dependence of the superspace lagrangian on $y$ and $\partial_y$. 
By expanding the superfields as $S(x,y,\theta)=\sum_n S_n(x,y)\theta^n$, we can identify $S_n(x,y)$ with the local 
5D fields. However our global supersymmetry knows little about the local 5D geometry, so that in general
the $S_n$'s are not normalized in a way that makes 5D covariance manifest. This is obviously not a problem: the 
correct normalization (as well as the correct covariant derivative structure) can always be obtained by local 
redefinitions of the $S_n$'s by powers of the warp factor \cite{Marti:2001iw,Hirayama:2003kk}. 
We can figure out the right rescaling that defines the canonical fields by considering the normalization 
of the supercharge. By eq.~(\ref{global}), $Q_1^L$ is related to the global supercharge
by $Q_1^L=e^{\sigma/2} Q^G$. Then by defining a new ``local'' superspace coordinate 
$\tilde \theta =e^{-\sigma/2}\theta$, and substituting the flat vielbein $\delta^\mu_a$ by the curved one 
$e_a^{~\mu}=e^\sigma \delta_a^\mu$, we can write
\be
Q_{1 \alpha}^L = \frac{\partial}{\partial \tilde\theta^\alpha} 
- i \sigma^a_{\alpha \alphadot} \bar {\tilde \theta}^{\dot \alpha} e_a^{~\mu} \partial_\mu \;.
\label{local} 
\ee
Not surprisingly, the presence of the vielbein shows that $Q_{1\alpha}^L$ is covariant and realizes the local 
supersymmetry subalgebra
\begin{equation}
\label{eq:ads5susyalgebra}
\left\{Q^L_{1\alpha}, \bar{Q}^L_{1 \alphadot} \right\} = -2 i \sigma^a_{\alpha \alphadot} 
e_a^{~m} \partial_m .
\end{equation}
Therefore, if we parametrize our superfields in terms of the ``local'' $\tilde \theta$ instead of the 
``global'' $\theta$, the field coefficients should correspond to the local canonical fields. From 
the definition $S = \sum_n  S_n \theta^n = \sum_n \tilde S_n \tilde \theta^n$ we conclude that the 
canonical fields $\tilde S_n$ must be defined as\footnote{To be precise, one should remember that 
the superfields coefficients $S_n$ involve in some cases 4D derivatives $\partial_\mu$. However,
this does not affect our conclusions, because the change $\theta \to \theta^\prime = e^{- \sigma/2}\theta$ 
in coordinates corresponds to the change $\partial_\mu \to \partial_a = e_a^{~\mu} \partial_\mu$
in derivatives. For instance, in the case of a chiral superfield we have 
\be
\Phi = e^{-i\bar \theta \sigma^a\bar \theta\delta_a^\mu\partial_\mu}
\bigl(\varphi+\chi\theta+F\theta^2\bigr )
= e^{-i\bar \tilde\theta \sigma^a\bar {\tilde\theta}e_a^\mu\partial_\mu}
\bigl(\varphi+\chi e^{\sigma/2}\tilde\theta+ e^\sigma F\tilde\theta^2\bigr ) \;.
\ee
}
\be
\tilde S_n(x,y)= e^{n\sigma /2} S_n(x,y) \;.
\ee
At first sight it might seem better to work directly with the $\tilde \theta$ coordinates. However, 
$Q_{1\alpha}^L$ does not commute with $\partial_y$, since by eq.~(\ref{global}) it depends 
explicitly on $y$, and therefore $\partial_y S$ is  not a superfield over the superspace defined 
by $\tilde \theta$. We could define a supercovariant $D_y$ derivative, but we find it more convenient 
to work with global, flat superspace.

Now that we know that it is possible to write the desired Lagrangian in 
term of standard $N=1$ superfields, we need to examine the gauge symmetry that this 
Lagrangian should possess. We parametrize the fluctuations around the background as
\begin{equation}
\label{eq:adsbckgnd}
ds^2 = e^{-2 \sigma} \left(\eta_{m n} + h_{mn}\right) dx^m dx^n 
+ 2 e^{- \sigma} h_{m y} d x^m dy + (1+h_{yy}) dy^2 \;,
\end{equation}
The linearized general coordinate transformations are then given by:
\begin{eqnarray}
\delta h_{mn} \a=\a \partial_m \xi_n + \partial_n \xi_m - 2 \sigma^\prime \eta_{m n} \xi_y 
\label{eq:hMNtransf} \;, \\
\delta h_{my} \a=\a e^{-\sigma} \partial_y \xi_m + e^\sigma \partial_m \xi_y \;, \\
\delta h_{yy} \a=\a \partial_y \xi_y \;.
\end{eqnarray}
Comparing with the flat case, we see that the warping is responsible for a new term 
proportional to $\sigma^\prime$ in the transformation law for $h_{mn}$. 

The embedding of component fields into superfields can be done as in the flat case, 
except that we need to introduce in this case a real prepotential $P_{\mathcal{T}}$ 
for $\mathcal{T}$ as well, in such a way that $\mathcal{T} = -1/4 \bar D^2 P_{\mathcal{T}}$. 
The transformation laws can then be written in terms of superfields, after introducing 
a prepotential $P_\Omega$ also for $\Omega$, as:
\begin{eqnarray}
\delta V_m \a=\a -\frac{1}{2} \sigmabar_m^{\alphadot \alpha} 
\left(\bar{D}_\alphadot L_\alpha - D_\alpha \bar{L}_\alphadot \right) 
\label{eq:transfADSV} \;, \\[-1.5mm]
\delta \Psi_\alpha \a=\a e^{- \sigma} \partial_y L_\alpha 
- \frac{1}{4} e^{\sigma} D_\alpha \Omega 
\label{eq:transfADSPsi} \;, \\[0.5mm]
\delta P_\Sigma \a=\a D^\alpha L_\alpha - 3 \sigma^\prime P_\Omega 
\label{eq:transfADSPSigma} \;, \\[1mm]
\delta P_{\mathcal{T}} \a=\a \partial_y P_\Omega 
\label{eq:transfADSPT} \;.
\end{eqnarray}
From the last two expressions, it follows that:
\begin{eqnarray}
\delta \Sigma \a=\a -\frac{1}{4}\bar{D}^2 D^\alpha L_\alpha - 3 \sigma^\prime \Omega 
\label{eq:transfADSSigma} \;, \\
\delta \mathcal{T} \a=\a \partial_y \Omega 
\label{eq:transfADST} \;.
\end{eqnarray}
Note that the new term proportional to $\sigma^\prime$ in the transformation law for 
$h_{mn}$ is encoded in superfield language in a new term in the transformation law for 
$\Sigma$. This is possible because, in addition to general coordinate invariance, 
the transformations parametrized by $L_\alpha$ also include Weyl and axial transformations. 
These extra conformal transformations can be fixed by setting the lowest component $s$ of 
the compensator $\Sigma$ to $0$. But the subgroup of transformations that preserve this 
gauge choice involves scale transformations that are correlated with diffeomorphims
and induce the appropriate extra term in eq.~\eqref{eq:hMNtransf}. To show this more 
precisely, let us consider the transformation laws of $h_{mn}$ and $s$ under diffeomorphisms 
with real parameters $\xi_M$ and complexified Weyl plus axial transformations with complex
parameter $\lambda$, as implied by eqs.~\eqref{eq:transfADSV} and \eqref{eq:transfADSSigma}:
\begin{eqnarray}
\delta h_{mn} \a=\a\partial_m \xi_n + \partial_n \xi_m
- \frac{2}{3} \partial_m \xi^m + \frac{1}{6} \eta_{mn} \left(\lambda + \lambda^* \right)
\label{transhmn} \;, \\
\delta s \a=\a 2 \partial_m \xi^m - 6 \sigma^\prime \xi_y - \lambda 
\label{transs} \;.
\end{eqnarray}
As anticipated we can now use the Weyl and axial symmetries associated to $\lambda$ to set 
$s$ to $0$. To preserve that gauge choice, however, diffeomorphisms must then be accompanied
by a suitable Weyl transformations with parameter
\begin{equation}
\lambda + \lambda^* = 4 \partial_m \xi^m - 12 \sigma^\prime \xi_y \;.
\end{equation}
Plugging this expression back into \eqref{transhmn}, we find that the net transformation 
law of the graviton under a diffeomorphism, after the conformal gauge-fixing, reproduces indeed 
eq.~\eqref{eq:hMNtransf}.

It is now straightforward to construct a superfield Lagrangian that is invariant under 
the transformations \eqref{eq:transfADSV}--\eqref{eq:transfADST} and reduces to 
\eqref{eq:linflatsugra} in the flat limit. It has the following expression:
\bea
\label{eq:bulklag}
\mathcal{L} = M_5^3 \! \int \! d^4 \theta \, e^{-2 \sigma} \a\!\bigg\{\!\a 
\!- V_{3/2}^m \big(\square + e^{2\sigma} \partial_y e^{-4\sigma} \partial_y \big) V_{3/2 m} 
- \frac 23 \Big[\partial_m V_0^m - \frac i2 \big(\Sigma - \Sigma^\dagger\big) \Big]^2 \nn \\
\a\!\!\a\!\! - \frac 12 \Big[e^{-\sigma} \partial_y \big(V_{0}^{\alphadot \alpha} 
+ V_{1/2}^{\alphadot \alpha} + V_{1}^{\alphadot \alpha} \big)
- \big(\bar{D}_\alphadot \Psi_\alpha - D_\alpha \bar{\Psi}_\alphadot \big)\Big]^2 \nn \\
\a\!\!\a\!\! + \frac{1}{4} \Big[e^{-\sigma}\big(\partial_y P_\Sigma 
+ 3 \sigma^\prime P_{\mathcal{T}} \big)
- \big(D^\alpha \Psi_\alpha + \bar{D}_\alphadot \bar{\Psi}^\alphadot \big) \Big]^2 \nn \\
\a\!\!\a\!\! +\, i \big(\mathcal{T} - \mathcal{T}^\dagger\big) \Big[\partial_m V_0^m 
- \frac i2 \big(\Sigma - \Sigma^\dagger\big) \Big] \bigg\} \;.
\eea
There is one important remark to make about this Lagrangian: it possesses an extra (accidental) local invariance
in addition to those we employed to derive it. Its parameter is a chiral superfield $W_\alpha$ and the transformation
laws are given by
\bea
\label{eq:newsymm}
\delta \Psi_\alpha \a=\a 3\sigma' e^{-\sigma} W_\alpha \;, \\
\delta P_\T \a=\a D^\alpha W_\alpha+\bar{D}_{\dot\alpha}\bar{W}^{\dot\alpha} \;.
\eea
This new symmetry is associated to a redundancy in the parametrization of $\T$ by
a prepotential $P_\T$. Indeed $P_\T$ shifts by a linear multiplet so that using $W_\alpha$ we can 
gauge away the newly introduced components of $P_\T$. By this invariance of the quadratic action, 
some combination of fields have no kinetic term. In order for our expansion to make sense when 
going to non-linear order, it is important to demand full invariance under this new transformation.

The component form of the Lagrangian (\ref{eq:bulklag}) is reported in appendix \ref{ap:components}. 
It reproduces the correct component Lagrangian for supergravity on a slice of $\text{AdS}_5$ at the 
linearized level.  The well known derivation of the low energy effective theory and KK decomposition 
then follows. It is however instructive to derive these results in terms of superfields.

\subsection{KK mode decomposition}
\label{KKdecomposition}

Let us start by constructing the zero mode superfield action. First of all, $\Psi_\alpha$ is $\Z_2$ odd 
and does therefore not have zero modes. Second, notice that if we choose $V^m(x,y) \to {\hat V}^m(x)$,
the mass terms, involving $\partial_y$ cancel out. On the other hand, the mass term involving $P_\Sigma$ 
and $P_\mathcal{T}$ in the third line is non-vanishing for $y$ independent field configurations.
One possible parametrization of the zero modes for which this term vanishes altogether\footnote{One 
can check that, by using the gauge freedom associated to $L_\alpha$ and $W_\alpha$, ${\hat P}_\T(x)$ 
and ${\hat P}_\Sigma(x)$ can be chosen to be purely chiral + antichiral (i.e no linear superfield component) 
while keeping $\Psi_\alpha=0$.} is defined by
\bea\label{eq:scalarmodes}
P_{\mathcal{T}}(x,y)\a\to\a {\hat P}_{\mathcal{T}}(x)\;, \phantom{-3\sigma(y) 
{\hat P}_\mathcal{T}(x)}\;\quad \mathcal{T}(x,y)\to\hat{\mathcal{T}}(x)\;, \nn \\
P_\Sigma(x,y)\a\to\a \hat P_\Sigma(x)-3\sigma(y) {\hat P}_\mathcal{T}(x)\;,\quad 
\Sigma(x,y)\to\hat \Sigma(x)-3\sigma(y)\hat {\mathcal{T}} \;.
\label{zeromodes}
\eea
Notice that the conformal compensator $\Sigma$ depends on $y$ precisely like the conformal factor 
of the metric does in the zero mode parametrization for the bosonic RS1 model \cite{Goldberger:1999un}.
In order to write the effective action in compact form it is useful to rewrite the superspin $0$ 
component as $\partial_mV_0^m=\chi +\chi^\dagger$ and  form the two combinations
\be
\label{sigmaAB}
\Sigma_A\equiv \Sigma +2i\chi \;,\quad \Sigma_B\equiv \Sigma -2i\chi \;.
\ee
It is important to notice that $\Sigma_A$ is gauge-invariant but $\Sigma_B$ is not.
The zero mode action, with the second and third lines in eq.~(\ref{eq:bulklag})
vanishing, depends  only on $\Sigma_A$. Defining the zero modes of the latter 
in analogy with eq.~(\ref{zeromodes}), i.e.
\be
\Sigma_A(x,y)\to \hat \Sigma(x)-3\sigma(y)\hat {\mathcal{T}}(x) +2i\hat \chi(x)
= \hat\Sigma_A(x)-3\sigma(y)\hat{ \mathcal{T}}(x) \;,
\ee
the 5D action for the zero modes can be rewritten simply as:
\bea
\label{eq:bulklagzm}
\mathcal{L} \to \frac {M_5^3}{\sigma^\prime} \! \int \! d^4 \theta \,\partial_y \a\!\bigg\{\!\a 
\!e^{-2\sigma} \Big[\frac 12 \hat V_{3/2}^m \square \hat V_{3/2 m} 
+ \frac 16 \bigl(\hat \Sigma_A^\dagger-3\sigma \hat{\mathcal {T}}^\dagger\bigr)
\bigl(\hat \Sigma_A-3\sigma \hat \T\bigr) \Big] \bigg\} \;.\quad
\eea
showing that the geometry of the boundaries is what matters in the low energy effective action.
The quadratic action for the zero modes is then
\bea
\label{effectivelinaction}
\mathcal{L}_{\rm eff} = \frac{M_5^3}{k} \!\int\! d^4 \theta \a\!\bigg\{\!\a 
\!\! -\Big(1-e^{-2 \pi k R}\Big) \hat V_{3/2}^m \square \hat V_{3/2 m} \nn \\
\a\!\!\a\!\! - \frac 13 \Big[\hat\Sigma_A^\dagger\hat\Sigma_A
-e^{-2\pi k R}\bigl (\hat\Sigma_A^\dagger-3\pi k R \hat \T^\dagger\bigr )
\bigl (\hat\Sigma_A-3\pi k R \hat \T\bigr ) \Big ]\;.\quad 
\eea
By expliciting the dependence of $\hat \Sigma_A$ on $\hat V_m$ it can be verified that this Lagrangian
is local as it should. Moreover, with the identification $\phi = \exp\,(\hat \Sigma/3)$ 
and $T = \pi R (1 + \hat{\mathcal{T}})$, it agrees with the quadratic expansion of 
the full non-linear result, which was inferred by general arguments in ref.~\cite{Luty:2000ec,Bagger:2000eh}:
\begin{equation}
\label{effectiveradionaction}
\mathcal{L}_{\rm eff} = -3 \frac {M_5^3}k \int d^4 \theta \,
\big(1 - e^{-k(T + T^\dagger)}\big) \phi^\dagger \phi \;.
\end{equation}

In fact, even though eq.~\eqref{effectivelinaction} is only valid at quadratic level in the superfield 
$\mathcal{T}$, the full non linear dependence on the radion superfield $T$ of eq.~\eqref{effectiveradionaction} 
can be deduced through the following argument.  At the linearized level, the real part of the scalar component $t$ of $\mathcal{T}$
coincides in principle with $h_{yy}\equiv \sqrt {g_{yy}}-1$ only up to higher order terms. However, one 
can argue that by a holomorphic field redefinition $\T\to\T^\prime=f(T)$ it should always be possible to 
choose ${\rm Re}\,t^\prime = \sqrt {g_{yy}}-1$ \cite{Luty:2000ec}.
Let us then choose $\T$ such that ${\rm Re}\,t=\sqrt{g_{yy}}-1$. Now, focussing on the constant mode
of $\sqrt{g_{yy}} $, we know that the low energy Lagrangian can only depend on it via the covariant
combination $\pi R\sqrt {g_{yy}}$ equalling the physical length of the fifth dimension. This is not yet enough to 
fully fix the dependence on $\T$. We need to use the constraints on the dependence on ${\rm Im }\,t \propto A_y$
the graviphoton fifth component. At tree level, The VEV of the low energy kinetic function corresponds to the effective 4D
Planck scale of the usual RS1 model, which does not depend on other bulk fields than the radion. In particular
it does not dependent on gauge fields like the graviphoton. This fixes completely the dependence on $\T$
to be obtained by the simple substitution $2\pi R\to \pi R(2+\hat\T+\T^\dagger)$, compatibly with our results. 
The dependence on $A_y$ is actually constrained even at the quantum level by the presence of an accidental 
(gauge) symmetry $A_y\to A_y + {\rm const.}$ of minimal 5D supergravity on $S_1/\Z_2$. The point is that the 
graviphoton appears in covariant derivatives via a $\Z_2$-odd charge, $\partial_y \to \partial_y +iq\epsilon(y) A_y$: 
basically the graviphoton is a $\Z_2$ odd gauge field used to gauge an even symmetry. It is then evident that 
a constant $A_y=a$ configuration can be gauged away by a gauge rotation with parameter 
$\alpha(y)=a\,\sigma(y)/(\pi k R)$.  Since we will be using the quadratic supergravity Lagrangian for the computation 
of the quantum effective action, we will later need this argument to turn the dependence of our result on $R$ into a 
dependence on the superfield $T$. 

The KK spectrum can be studied in a similar way. For doing so, it is convenient to work in the gauge 
$\Psi_\alpha=0$, where it is evident that $V_{1/2}$ and $V_1$ do not propagate while $V_{3/2}$ has 
the KK decomposition of the graviton in RS1. We then exploit the local invariances associated to the 
parameters $P_\Omega$, $W_\alpha$ and $L_\alpha$ to bring the fields $P_\T$ and $P_\Sigma$ in 
a convenient form. In addition to a chiral part of superspin $0$, these real superfields contain a linear 
superfield mode of superspin $1/2$ that does not propagate, as will now argue.
First, by using $P_\Omega$ and $W_\alpha$, we can always bring $P_\T$ in the form $e^{2 \sigma} P_{\bar \T}(x)$,
where $P_{\bar \T}$  is $y$-independent and satisfies $\bar{D}^2D^\alpha P_{\bar \T}=0$. This choice eliminates all the modes 
in $P_\T$ apart from a chiral radion zero mode. Next, by using the residual freedom $L_\alpha = S_\alpha$ 
with $S_\alpha(x)$ chiral and constant over $y$ we can eliminate the $y$ independent linear superfield mode in 
$P_\Sigma$. Note that $V_m$ and $\Psi_\alpha$ are unaffected by these transformations. In this way, in the 
superspin $1/2$ sector of $P_\Sigma$ and 
$P_\T$, only the non-trivial KK modes of $P_\Sigma$ are left. Like for $V_{1/2}$ and $V_1$, their kinetic 
Lagrangian is a simple quadratic term proportional to $(\partial_y P_\Sigma)^2$ with a trivial mass shell 
condition setting the modes themselves to zero. In what follows, we can then concentrate on the components
with superspin $0$ and work directly with the chiral fields $\T$ and $\Sigma$. Compatibly with the above 
discussion, the decompositions of the fields are chosen as follows:
\bea
\T(x,y) \a=\a e^{2\sigma(y)}\bar \T(x) \;, \\
\Sigma(x,y) \a=\a -\frac{3}{2}e^{2\sigma(y)}\bar\T(x)+\tilde \Sigma(x,y) \;.\qquad 
\eea
The corresponding expressions for $\Sigma_{A,B}$ (cfr. eq.~(\ref{sigmaAB})) are then given by
\be
\label{tildesigma}
\Sigma_A(x,y)=-\frac{3}{2}e^{2\sigma}\bar\T(x)+\tilde \Sigma_A(x,y)\;,\;\;
\Sigma_B(x,y)=-\frac{3}{2}e^{2\sigma}\bar\T(x)+\tilde \Sigma_B(x,y) \;.
\ee
With this parametrization the Lagrangian for the chiral fields becomes
\be
\label{kineticdiagonal}
{\cal L}_{\bar \T,\tilde\Sigma}  =  \int \!d^4\theta\left \{\frac{3}{4} e^{2\sigma} \bar \T^\dagger \bar \T 
- \frac{1}{3} e^{-2\sigma} \tilde \Sigma_A^\dagger \tilde \Sigma_A 
+ \frac 14 e^{-4\sigma} \bigl (\partial_y \tilde \Sigma_B^\dagger \frac{1}{\square}\partial_y\tilde \Sigma_A
+{\rm h.c.}\bigr )\right \} \;.
\ee
It is useful, and straightforward, to decompose $\tilde \Sigma_A$ and $\tilde \Sigma_B$ in a complete set 
of orthogonal KK modes satisfying:
\begin{equation}
\partial_y e^{-4 \sigma} \partial_y \tilde \Sigma_{A,B}^{(n)} =m_n^2 e^{-2 \sigma} \tilde \Sigma_{A,B}^{(n)} \;.
\end{equation}
The KK modes Lagrangian can then be written (with a convenient normalization) as:
\bea
\label{kineticdiagonalKK}
{\cal L}_{\bar \T,\tilde\Sigma^{(n)}} \a=\a  \int \!d^4\theta\bigg \{\frac{3}{4}\left(e^{2 \pi R}-1 \right) \bar \T^\dagger \bar \T \\
\a\;\a \hspace{37pt} -\frac{1}{3} \left(1-e^{-2 \pi R} \right) \sum_n \left[
\tilde \Sigma_A^{\dagger (n)} {\tilde \Sigma_A }^{(n)}
+ m_n^2\left( \tilde \Sigma_B^{\dagger (n)} \frac{1}{\square} {\tilde \Sigma_A}^{(n)} 
+{\rm h.c.}\right) \right] \bigg\} \;. \nn
\eea
This parametrization makes it manifest that the non-zero KK modes of $\tilde \Sigma_B$ act as Lagrange 
multipliers for the modes of $\tilde \Sigma_A$ \cite{Gregoire:2004ic}. The only physical modes that are left are therefore $\bar\T$ 
along with $\tilde\Sigma_A^{(0)}(x)$: they represent an alternative parametrization of the zero modes, 
in which the radion does not mix kinetically with the 4D graviton. This is the superfield analogue of the radion 
parametrization discussed in ref.~\cite{Charmousis:1999rg}. 

\subsection{Boundary Lagrangians}

The only interactions that are needed for our calculation are those between brane and bulk fields. At the fixed points, 
$\Psi_\alpha$ vanishes, $V_m$ and $\Sigma$ undergo the same transformations of linearized 4D supergravity 
(remember that $\Omega$ is odd and vanishes at the boundaries) and finally $\T$ is the only field transforming 
under $\Omega$. By this last property, $\T$ cannot couple to the boundary, so that only $V_m$ and $\Sigma$ 
can couple and they must do so precisely like they do in 4D supergravity (see \cite{Buchbinder:2003qu}). What is left
are the 4D superdiffeomorphisms of the boundaries. The presence of warping shows up in the boundary Lagrangian 
via the suitable powers of the warp factor. These are easy to evaluate according to our discussion in the previous 
section. Using locally inertial coordinates $\tilde x$ and $\tilde \theta$, no power of the warp factor should appear 
in the invariant volume element $d^4\tilde x d^4 \tilde\theta$  for the K\"ahler potential and $ d^4\tilde x d^2 \tilde \theta$ 
for the superpotential. From the ordinary RS1 model, we know that $\tilde x=e^{-\sigma}x$, where $x$ are the global 
coordinates, while from the previous section we have learned that $\tilde \theta= e^{-\sigma/2} \theta$. 
We conclude that the warp factors multiplying the K\"ahler and the superpotential of the actions localized at 
$y=y_i$ must be equal respectively to $e^{-2\sigma(y_i)}$ and $e^{-3\sigma(y_i)}$.

Let us consider a chiral superfield $\Phi_i$ localized at the brane at $y_i$ and with quadratic K\"ahler 
potential $\Omega_i = \Phi_i^\dagger \Phi_i$. As we already argued, the couplings of $\Phi_i$ to
$V_m$ and $\Sigma$ are the same as in ordinary 4D supergravity.
For our purposes, since we are only interested in the 1-loop K\"ahler potential,  it is sufficient to consider terms 
that are at most quadratic in $V_m$ and $\Sigma$, with derivatives acting on at most one of  $\Phi_i$ and $\Phi^\dagger_i$. 
The relevant part of the 4D boundary Lagrangian at $y_i$ is then given by
\bea
\label{eq:interactionlag}
\mathcal{L}_{i} \a=\a \int \! d^4 \theta e^{-2\sigma(y_i)} \Big[\Phi_i^\dagger \Phi_i
\Big(1+\frac 13 \Sigma^\dagger \Big) \Big(1 + \frac 13 \Sigma\Big) 
+ \frac{2}{3}i \Phi_i^\dagger \overleftrightarrow{\partial_m} \Phi_i V^m \nn \\
\a\;\a \hspace{70pt} -\, \frac{1}{6} \Phi_i^\dagger \Phi_i V^m K_{mn} V_n \Big]\;.
\eea
To compute the effective K\"ahler potential, we split the matter field into classical background $\bar \Phi_i$ and 
quantum fluctuation $\pi_i$: $\Phi_i= \bar \Phi_i+\pi_i$.  We then rewrite the Lagrangian in term of the different superspin 
components defined previously\footnote{Note that these projectors are ill-defined when $p^2=0$. However, since 
we are interested in a loop computation with non-vanishing virtual momentum, this doesn't cause any problem.}
\bea
\label{eq:interactionlagproj}
\mathcal{L}_{i} = \int \! d^4 \theta e^{-2\sigma(y_i)}\a\!\Big[\!\a\!\!\bar\Phi_i^\dagger \bar \Phi_i+\pi_i^\dagger\pi_i
+\frac{1}{3}\bar\Phi_i^\dagger \pi_i \Sigma_A^\dagger +\frac{1}{3}\bar\Phi_i \pi_i^\dagger \Sigma_A \nn \\
\a\!\!\a\!\!- \frac{1}{3} \bar\Phi_i^\dagger \bar\Phi_i \Big(V_{3/2}^m \square V_{3/2 m} 
- \frac{1}{3}\Sigma_A^\dagger\Sigma_A \Big)\Big]\;.
\eea
In this expression, we have neglected terms involving derivative of the background and interactions among the 
quantum fluctuations, as they do not do not affect the K\"ahler potential at 1-loop. Notice that the Lagrangian 
involves only the gauge invariant quantities $\Sigma_A$ and $V_{3/2}^m$. This property will be very useful for 
the computation of the next subsection. The effect of localized kinetic 
terms is easily obtained by generalizing the localized K\"ahler potential as $\Omega_i \to \Omega_i - 3 M_i^2$. In the 
above Lagrangian, this amounts to change $\bar \Phi^\dagger \bar \Phi \to \bar \Phi^\dagger\bar \Phi -3M_i^2$
in those terms that are quadratic in the background, without affecting the terms that are instead linear in the 
background and involve fluctuations.

Finally, by using our parametrization of the radion and compensator zero modes, we can check that the boundary 
contribution to the low energy effective Lagrangian is just the linearized version of the full non-linear result,
i.e.:
\be
\label{effectivematteraction}
\mathcal{L}_{\rm eff} =\int \! d^4 \theta \Big\{\Omega_0(\Phi_0,\Phi_0^\dagger) + \Omega_1(\Phi_1,\Phi_1^\dagger)
e^{-k(T+T^\dagger)}\Big\} \phi^\dagger\phi \;.
\ee

\subsection{One loop effective K\"ahler potential}

We now have all the necessary ingredients to set up more concretely the calculation of gravitational quantum corrections
to the 4D effective K\"ahler potential. 
As we restrict our attention to the effective K\"ahler potential, we neglect all derivatives
on the external fields. The supergraph calculation that needs to be done becomes then very similar to the 
calculation of the Coleman--Weinberg potential in a non-supersymmetric theory \cite{Coleman:1973jx}.
Similar superfield computations have already been done for the gauge corrections to the K\"ahler potential 
in 4D supersymmetric theories in \cite{Grisaru:1996ve}. 

Normally, the most convenient procedure for doing this kind of computation is to add a suitable gauge-fixing term 
to the Lagrangian and work in generalized $R_\xi$ gauges. In the case at hand, however, it turns out that there 
is a much simpler approach. The point is that the fields $V_{3/2}$ and $\Sigma_A$ appearing in the boundary 
Lagrangian are already gauge-invariant combinations. Their propagator is gauge-independent and so we can work 
directly in the unitary gauge (defined in section \ref{KKdecomposition}), where the bulk Lagrangian takes it simplest form. 
Let us first examine the contribution from the $\Sigma_A$ field. It couples to matter
at the boundaries, and the quantity that is relevant for the computation is the propagator 
$\langle \Sigma_A(x,y_i)^\dagger\Sigma_A(x,y_j)\rangle$ connecting two branes at positions $y_i$ and $y_j$.
The relevant Lagrangian for computing the propagator has already been worked out and is given by eq.~(\ref{kineticdiagonalKK}). 
For the non-zero KK modes, we can write the kinetic matrix in two by two matrix notation as:
\be
K_{\tilde \Sigma_A^{(n)}, \tilde \Sigma_B^{(n)}} = \left(\begin{matrix} 
\displaystyle{-\frac{1}{3}} & \displaystyle{\frac{m_n^2}{\square}} \medskip\ \\
\displaystyle{\frac{1}{4}\frac{m_n^2}{\square}} & 0
\end{matrix}\right) \;.
\ee
Inverting this matrix, we find that  for $m_n\neq 0$ we have $\langle \tilde \Sigma_A(x)^{\dagger (n)} \tilde \Sigma_A(x)^{(n)}\rangle =0$, 
and correspondingly, these modes do not contribute to $\langle \Sigma_A(x,y_i)^\dagger\Sigma_A(x,y_j)\rangle=0$  (c.f. eq. \eqref{tildesigma}). The only modes 
that are left are thus the zero modes $\bar \T$ and  $\tilde \Sigma_A^{(0)}(x)$. The kinetic operator for these two fields is, again 
in two by two matrix notation:
\be
K_{\tilde \Sigma_A^{(0)}, \bar \T} = \left(\begin{matrix} 
\displaystyle{-\frac 13\Big(1 - e^{-2\pi R}\Big)} & 0 \medskip\ \\
0 & \displaystyle{\frac 34 \Big(e^{2\pi R} - 1 \Big)}
\end{matrix}\right) \;.
\ee
It is straightforward to verify that the resulting propagators for $\tilde \Sigma_A^{(0)}$ and $ \mathcal{\tilde T}$  give no
contribution to $\langle \Sigma_A(x,y_i)^\dagger\Sigma_A(x,y_j)\rangle$. This is only possible thanks to the 
ghostly nature of the $\tilde \Sigma_A^{(0)}$ field, which contains indeed the conformal mode of the 4D graviton. The 
cancellation is spoiled if $y_i = y_j$, and in that case one gets a non-trivial contribution to the propagator. However, the 
volume dependence is such that  it contributes only to local terms, we therefore conclude that the zero modes are also 
irrelevant for our calculation. Summarizing, the above reasoning shows that the whole $\Sigma_A$ field is not relevant to 
our computation. We  therefore need to focus only on $V^m_{3/2}$, and the relevant Lagrangian has the form
\be
\label{eq:fulllagrangian32}
\mathcal{L}_{V_{3/2}} = - M_5^3 \int d^4 \theta\, V_m^{3/2}  \Big[
e^{-2\sigma} \Big(1 - \rho_0 \delta_0(y) - \rho_1 \delta_1(y) \Big) \square
+ \partial_y e^{-4\sigma} \partial_y \Big] V^m_{3/2} \;.
\ee
The quantities $\rho_{i}$ represent the boundary corrections to kinetic terms\footnote{Note that we have defined 
these quantities with an overall negative sign, as in ref.~\cite{Rattazzi:2003rj}.}, including the effect of matter field VEVs. 
For phenomenological application, we will be interested in the case $\rho_i=(-3M_i^2+\Phi_i^\dagger\Phi_i)/(3 M_5^3)$. 
Inverting the above kinetic term the superspin structure factors out in an overall projector:
\begin{equation}
\label{eq:propV3/2}
\langle V_m^{3/2}\left(x_1,y_1,\theta_1\right) V_n^{3/2}\left(x_2,y_2,\theta_2\right)\rangle = 
- \Pi_{3/2}^{mn}\, \delta^4 \! \left(\theta_1 - \theta_2\right) \Delta \left(x_1,y_1;x_2,y_2\right) \;,
\end{equation}
where $\Delta$ is the propagator of a real scalar field in a slice of $\text{AdS}_5$, in presence
of boundary kinetic terms, defined by the equation
\bea
\label{eq:scalarprop}
\Big[e^{-2 \sigma} \Big(1 \!-\! \sum_i \rho_i \delta_i(y) \Big) \square
+\! \partial_y e^{-4 \sigma} \partial_y\Big]  \Delta \left(x_1,y_1;x_2,y_2\right) 
= \delta^4(x_1-x_2) \delta(y_1-y_2) \;.\;\;
\eea
This is compatible with the results found in ref. \cite{Rattazzi:2003rj} for the flat case. 
By applying the methods of refs.~\cite{Grisaru:1996ve}, the 1-loop effective action can be written 
in a factorized form as\footnote{This relation can we obtained by first defining the trivial 
Gaussian functional integral for a non-propagating field as
\be
\int {\cal D} V_m e^{i\int \!d^4x d^4\theta V_m V^m} = 1 \;.
\ee
The interesting case is then obtained by computing a similar functional integral where the trivial kinetic function 
$\eta_{m n}$ is substituted with $\eta_{mn}+\Pi_{mn}^{3/2}(\Delta^{-1}-1)$. The resummation in perturbation theory 
of all the insertions of $\Pi^{mn}_{3/2}$ leads to eq.~(\ref{full1loopaction}).}
\bea
\label{full1loopaction}
\Gamma \a=\a \frac{i}{2} \!\int\! d^4x \!\int\! d^4x^\prime \, \delta^4\! (x - x^\prime) 
\!\int\! dy \!\int\! dy^\prime \,\delta(y-y^\prime) 
\!\int\! d^4\theta \!\int\! d^4\theta^\prime \,\delta^4(\theta-\theta^ \prime) \nn \\
\a\;\a \Big[\eta_{m n} \Pi_{3/2}^{mn}\delta^4(\theta-\theta^\prime)(\ln \Delta^{-1})
(x,y,\theta;x^\prime,y^\prime,\theta^\prime) \Big] \;.
\eea
By using then the identity
\bea
\label{relation}
\int \! d^4\theta \! \int \! d^4 \theta^\prime\, \delta^4 (\theta - \theta^\prime)\,
\Big[\eta_{mn} \Pi_{3/2}^{mn}\, \delta^4(\theta - \theta^\prime)\Big] 
= \int d^4\theta \frac {-4}{\square} \;,
\eea
expanding the scalar propagator $\Delta$ in KK modes, and continuing to Euclidean momentum, 
the 1-loop correction to the K\"ahler potential
is finally found to be
\be
\label{kahlerfromscalar}
\Delta\Omega_{\rm 1-loop}=\frac{1}{2}\int \frac{d^4p}{(2\pi)^4}\sum_n \frac{-4}{p^2}\ln (p^2+\bar{m}_n^2) \;.
\ee
Apart from the $-4/p^2$ factor, arising from the superspace trace, this formula is just the Casimir energy
of a scalar whose propagation is governed by eq.~(\ref{eq:scalarprop}). From now on, we will therefore 
concentrate on this simple scalar Lagrangian.

It is worth noticing that the fact that the boundary interactions involve only the gauge invariant superspin $3/2$ and 
$0$ components is actually a general result valid for any number of extra dimensions. What is specific to 5D theories
is that the scalar component does not possess any propagating KK mode, and therefore cannot contribute to the genuinely 
calculable effects, like the Casimir energy or the radion-mediated and brane-to-brane-mediated soft masses.
In higher dimensions, this is no longer true, so that the scalar channel can in principle contribute. However it turns out that, 
for a simple but remarkable property of eq.~(\ref{eq:interactionlagproj}), this happens only for the Casimir energy but not for 
the terms that are quadratic in the matter fields. The reason is that  eq.~(\ref{eq:interactionlagproj}) inherits from the original 
quadratic matter Lagrangian a rescaling symmetry, thanks to which the compensator dependence can be fully reabsorbed by 
a redefinition of the matter field. In eq.~(\ref{eq:interactionlagproj}), this rescaling amounts to the shift 
$\pi_i\to \pi_i-\bar \Phi_i\Sigma_A/3$. Diagrammatically, the existence of this rescaling symmetry reveals itself in the mutual 
cancellations of the two diagrams shown in fig.~\ref{fig:cancel}. Using this property, we can, for diagrams involving matter, 
do our computation by first eliminating the compensator $\Sigma$, rather than $\Sigma_A$. After that we can work 
in a generalization of Landau's gauge, where the $\langle V^mV^n\rangle$ propagator is proportional to the 
$\Pi_{3/2}^{mn}$ projector. In this gauge, the diagrams involving cubic vertices cancel individually, in analogy with 
what happens for the calculation of the effective potential in non-supersymmetric theories.
\begin{figure}
\begin{center}
\begin{picture}(210,150)(0,30)
\DashArrowLine(20,100)(50,150){5}
\DashArrowLine(-10,150)(20,100){5}
\Photon(20,100)(50,50){4}{6}
\Photon(20,100)(-10,50){4}{6}
\put(-20,160){$ \bar \Phi_i$}
\put(50,160){$ \bar \Phi_i^\dagger$}
\put(-20,34){$V_0$}
\put(50,34){$V_0$}
\DashArrowLine(190,100)(200,150){5}
\DashArrowLine(140,150)(150,100){5}
\DashArrowLine(150,100)(190,100){5}
\put(170,90){$\pi_i$}
\Photon(190,100)(200,50){4}{6}
\Photon(150,100)(140,50){4}{6}
\put(130,160){$\bar \Phi_i$}
\put(200,160){$\bar \Phi_i^\dagger$}
\put(130,34){$V_0$}
\put(200,34){$V_0$}
\put(85,100){$+$}
\put(230,100){$= \quad 0$}
\end{picture}
\end{center}
\caption{\em Cancellation of two diagrams contributing to the effective K\"ahler potential.}
\label{fig:cancel}
\end{figure}
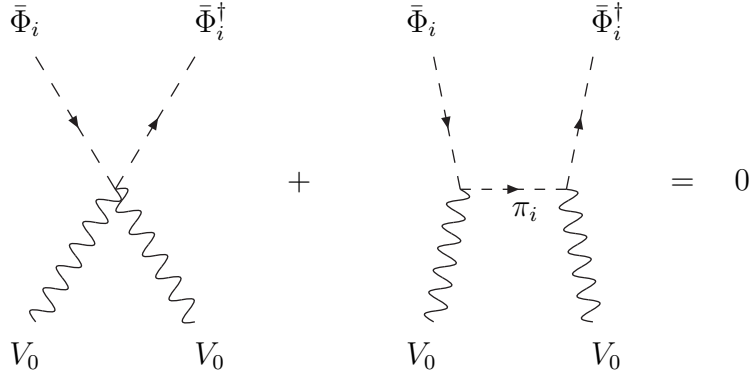
In order to properly define the gauge that is needed to implement the above program, we must add 
a suitable gauge-fixing to the Lagrangian. Since the gauge transformation 
$\delta V_{\alpha \alphadot} = D_\alpha \bar{L}_\alphadot - \bar{D}_\alphadot L_\alpha$ spans the 
subspace of components with superspin $0$, $1/2$ and $1$, the parameter $L_\alpha$ can be 
used to adjust the components $V_0^m$, $V_{1/2}^m$ and $V_1^m$, but not $V_{3/2}^m$. 
Correspondingly, the most general acceptable gauge-fixing Lagrangian consists of a combination of 
quadratic terms for $V_0^m$, $V_{1/2}^m$ and $V_1^m$. 
The choice that allows to reach the gauge where only $V_{3/2}^m$ propagates turns out to be given 
by the the following expression:
\begin{equation}
\label{eq:gaugefixingerm}
\mathcal{L}_{\text{gf}} = \int d^4 \theta e^{-2 \sigma} 
\Big[\!-\!\frac{1}{\xi} V_m \left(\eta^{mn} - \Pi_{3/2}^{mn} \right) \square V_n 
- \frac{2}{3} V_m \Pi_0^{mn} \square V_n \Big] \;.
\end{equation}
The part of the total Lagrangian that is quadratic in $V_m$ then becomes:
\bea
\label{eq:bulklaggaugefixed}
\mathcal{L}_{\text{quad}} = \int d^4 \theta \a\!\Big[\!\a
- V_m \Pi_{3/2}^{mn} \big(e^{-2\sigma} \square + \partial_y e^{-4\sigma} \partial_y \big) V_n \nn \\
\a\!\!\a - \frac 1\xi V_m \big(\eta^{mn} - \Pi_{3/2}^{mn} \big) e^{-2\sigma} \square V_n \Big] \;.
\eea
For $\xi \rightarrow 1$, and for a 4D theory where the extra superfields would be absent, this 
procedure would define the analogue of the super-Lorentz gauge. Its form coincides with the 
one that was used in ref.~\cite{Buchbinder:2003qu}, except for terms involving $\Sigma$,
$\mathcal{T}$ and $\Psi_\alpha$. For $\xi \rightarrow 0$, instead, it defines the analogue of the 
Landau gauge that we need. Indeed, it is clear that when $\xi$ is sent to $0$ only the $V_{3/2}^m$ 
component can propagate. Moreover, since $V_{3/2}^m$ does not couple to $\Psi_\alpha$, $\Sigma$ 
and $\mathcal{T}$, the full $\langle V_m V_n\rangle$ is now {\it exactly} given by the left hand side 
of eq.~(\ref{eq:propV3/2}).

\section{Explicit computation and results}\setcounter{equation}{0}
\label{sec:results}

As demonstrated in the last section, the full 1-loop correction to the K\"ahler potential is encoded 
in the spectrum of a single real 5D scalar field $\varphi$ with Lagrangian
\be
{\cal L} = \frac 12  e^{-2\sigma(y)} \Big[-(\partial_\mu \varphi)^2 - e^{-2\sigma(y)}(\partial_y \varphi)^2
+ \Big(\rho_0 \delta_0(y) + \rho_1 \delta_1(y) \Big) (\partial_\mu \varphi)^2 \Big] \;.
\label{Lagphi}
\ee
More precisely, see eq.~(\ref{kahlerfromscalar}), the effective 1-loop K\"ahler potential is obtained by 
inserting a factor $-4/p^2$ in the virtual momentum representation of the scalar Casimir energy. It is 
also understood that the circumference $2 \pi R$ should be promoted to the superfield $T + T^\dagger$
(see section \ref{KKdecomposition}), and similarly the constants $\rho_i$ should be promoted to the superfields 
$(- 3 M_i^2 + \Phi_i^\dagger \Phi_i )/(3 M_5^3)$. In the end, the superspace structure therefore only 
counts in a suitable way the various degrees of freedom via eq.~\eqref{relation}.
The factor $4$ takes into account the multiplicity of bosonic and fermionic degrees of freedom 
and the factor $1/p^2$ the fact that the K\"ahler potential determines the component effective action 
only after taking its $D$ component.

Let us now come to the computation. Along the lines of ref.~\cite{Rattazzi:2003rj}, we find it 
convenient to start from $\rho_{i}=0$ and to construct the full result by resuming the Feynman 
diagrams with all the insertions of $\rho_{i}$.  The building blocks for the computation of the 
matter-dependent terms are the boundary-to-boundary propagators connecting the points $y_i$ and 
$y_j$, with $y_{i,j}=0,\pi R$:
\be
\Delta_{ij}(p) = \sum_n e^{-\frac 32 \sigma(y_i)} e^{-\frac 32 \sigma(y_j)} \frac {\Psi_n(y_i) \Psi_n(y_j)}{p^2 + m_n^2}
=e^{-\frac 32 \sigma(y_i)} e^{-\frac 32 \sigma(y_j)}\Delta(p,y_i,y_j)  \;.
\label{defprop}
\ee
Here and in what follows, $\Psi_n(y)$ and $m_n$ denote the wave functions and the masses of the KK 
modes of the scalar field $\varphi$, in the limit $\rho_i=0$. The quantity $\Delta$ appearing in the second equality
is then by definition the propagator of the scalar field $\varphi$ in the same limit and in mixed momentum-position 
space: momentum space along the non-compact directions, and configuration space along the fifth. The 
exponential factors have been introduced for later convenience. The quantity that is relevant to compute the 
matter-independent term is instead the following spectral function:
\be
Z(p) = \prod_n \big(p^2 + m_n^2\big) \;.
\label{E}
\ee

The explicit expressions for the above quantities are most conveniently written in terms of the functions 
$\hat I_{1,2}$ and $\hat K_{1,2}$, defined in terms of the standard Bessel functions $I_{1,2}$ and $K_{1,2}$ 
as
\be
\hat I_{1,2}(x) = \sqrt{\frac {\pi}{2}} \sqrt{x}\,I_{1,2}(x) \;,\;\;
\hat K_{1,2}(x) = \sqrt{\frac {2}{\pi}} \sqrt{x}\,K_{1,2}(x) \;.
\label{hatIK}
\ee
These functions are elliptic generalizations of the standard trigonometric functions,
and satisfy the relation
\be
\hat I_1(x) \hat K_2(x) + \hat K_1(x) \hat I_2(x) = 1 \;.
\label{rel}
\ee
Their asymptotic behavior at large argument $x \gg 1$ is
\begin{equation}
\begin{array}{ll}
\displaystyle \hat K_{1}(x)  \simeq e^{-x}  \bigg[1 + \frac{3}{8x}-\frac{15}{128x^2}+\cdots\bigg]\;, & 
\displaystyle \hat I_1(x) \simeq \frac{e^x}{2}\bigg[1-\frac{3}{8x}-\frac{15}{128x^2}+\cdots\bigg]\;, \nn \\[5mm] 
\displaystyle \hat K_{2}(x)  \simeq e^{-x}  \bigg[1 + \frac{15}{8x}+\frac{105}{128x^2}+\cdots\bigg]\;,& 
\displaystyle \hat I_2(x) \simeq \frac{e^x}{2} \bigg[1-\frac{15}{8x}+\frac{105}{128x^2}+\cdots\bigg] \;.,\nn
\end{array}
\end{equation}
and their asymptotic behavior at small argument $x \ll 1$ is instead
\begin{equation}
\begin{array}{ll}
\displaystyle \hat K_1(x) \simeq \sqrt{\frac {2}{\pi}} \Big[x^{-\frac 12} + \cdots \Big] &
\displaystyle \hat I_1(x) \simeq \sqrt{\frac {\pi}{2}} \Big[\frac 12 x^{\frac 32} + \cdots \Big] \;, \nn \\[5mm] 
\displaystyle \hat K_2(x) \simeq \sqrt{\frac {2}{\pi}} \Big[2 x^{-\frac 32} + \cdots \Big]\;, &
\displaystyle \hat I_2(x) \simeq \sqrt{\frac {\pi}{2}} \Big[\frac 18 x^{\frac 52} + \cdots \Big] \;. \nn
\end{array}
\end{equation}

Consider first the computation of the quantities (\ref{defprop}).  Rather than computing them directly as 
infinite sums over KK mode masses, we derive them from particular cases of the propagator 
$\Delta(p,y,y^\prime)$ for $\varphi$, which is given by the solution with Neumann boundary conditions 
at $y$ equal to $0$ and $\pi R$ of the following differential equation:
\be
\Big(e^{-2ky}p^2 - \partial_y e^{-4ky} \partial_y \Big) \Delta(p,y,y^\prime) 
= \delta(y - y^\prime) \;.
\ee 
The solution of this equation is most easily found by switching to the conformal variable
$z = e^{ky}/k$, in which the metric is given by eq.~(\ref{confmetr}) and the positions of the 
two branes by $z_0 = 1/k$ and $z_1 = e^{k \pi R}/k$.
Notice that in these coordinates, a rescaling $z\to z\lambda$ is equivalent to a shift $z_{0,1}\to \lambda z_{0,1}$ 
of both boundaries plus a Weyl rescaling $g_{\mu\nu}\to g_{\mu\nu}/\lambda^2$ of the metric along 
the 4D slices. Therefore, both $1/z_0$ and $1/z_1$ have the properties of conformal compensators, 
and by locality it is then natural to identify them at the superfield level with the superconformal 
compensators at the respective boundaries: $(k z_0)^{-2}\to \phi\phi^\dagger$ and 
$(k z_1)^{-2} \to \phi\phi^\dagger e^{-k(T+T^\dagger)}$.
Defining also $u = {\rm min}(z,z^\prime)$ and $v = {\rm max}(z,z^\prime)$, the propagator is 
given by \cite{Gherghetta:2000kr,Randall:2001gb}:
\bea
\Delta(p,u,v) \a=\a\! \frac
{\Big[\hat I_1(p z_0) \hat K_2(p u) +\! \hat K_1(p z_0) \hat I_2(p u) \Big]\!
\Big[\hat I_1(p z_1) \hat K_2(p v) +\! \hat K_1(p z_1) \hat I_2(p v) \Big]}
{2p\, (k u)^{-\frac 32} (k v)^{- \frac 32}
\Big[\hat I_1(p z_1) \hat K_1(p z_0) - \hat K_1(p z_1) \hat I_1(p z_0) \Big]}
\,. \qquad\; \label{propgen}
\eea
The brane restrictions of the general propagator (\ref{propgen}) defining 
eqs.~(\ref{defprop}) are then easily computed. The factors 
$e^{-\frac 32 k y_{i,j}}$ that have been introduced cancel the factors 
$k z_{i,j}^{3/2}$ appearing in (\ref{propgen}). Moreover, one of the factors
in the numerator is always equal to 1 thanks to eq.~(\ref{rel}). The results finally
read
\bea
\Delta_{00}(p) \a=\a \frac {1}{2p} \frac
{\hat I_1(p z_1) \hat K_2(p z_0) + \hat K_1(p z_1) \hat I_2(p z_0)}
{\hat I_1(p z_1) \hat K_1(p z_0) - \hat K_1(p z_1) \hat I_1(p z_0)} \;, 
\label{prop00} \\
\Delta_{11}(p) \a=\a \frac {1}{2p} \frac
{\hat I_1(p z_0) \hat K_2(p z_1) + \hat K_1(p z_0) \hat I_2(p z_1)}
{\hat I_1(p z_1) \hat K_1(p z_0) - \hat K_1(p z_1) \hat I_1(p z_0)} \;, 
\label{prop11} \\[2mm]
\Delta_{01,10}(p) \a=\a \frac {1}{2p} \frac
{1}{\hat I_1(p z_1) \hat K_1(p z_0) - \hat K_1(p z_1) \hat I_1(p z_0)} \;.
\label{prop0110}
\eea

Next, consider the formal determinant (\ref{E}). Although this is not precisely a propagator, it is still a function 
of the spectrum and can be functionally related to the propagator in eq.~(\ref{propgen}). Indeed, the masses 
$m_n$ are defined by the positions of the poles $p = i m_n$ of (\ref{propgen}). These are determined by the 
vanishing of the denominator, that is by the equation
\be
F(i m_n) = \hat I_1(i m_n z_1) \hat K_1(i m_n z_0) - \hat K_1(i m_n z_1) \hat I_1(i m_n z_0) = 0 \;.
\label{spectrum}
\ee
The infinite product in eq.~(\ref{E}) is divergent. More precisely, it has the form of a constant divergent 
prefactor times a finite function of the momentum. In order to compute the latter, we consider the quantity 
$\partial_p\, {\rm ln} Z(p) = \sum_n 2p/(p^2 + m_n^2)$. The infinite sum over the eigenvalues, which are 
defined by the transcendental equation (\ref{spectrum}), is now convergent and can be computed with 
standard techniques, exploiting the so-called Sommerfeld--Watson transform. The result is simply given 
by $\partial_p\, {\rm ln} F(p)$. This implies that $Z(p) = F(p)$, up to the already mentioned  irrelevant 
infinite overall constant. Omitting the latter, we have therefore
\be
\label{eq:Z}
Z(p) = \hat I_1(p z_1) \hat K_1(p z_0) - \hat K_1(p z_1) \hat I_1(p z_0) \;.
\ee

\begin{figure}
\begin{center}
\begin{picture}(120,160)(0,0)
\PhotonArc(0,80)(40,0,360){4}{28}
\PhotonArc(0,80)(40,2,362){4}{28}
\DashArrowLine(0,120)(1,155){4}
\DashArrowLine(-40,80)(-75,80){4}
\DashArrowLine(40,80)(75,80){4}
\DashArrowLine(28.28,108.28)(53.02,133.02){4}
\DashArrowLine(-28.28,108.28)(-53.02,133.02){4}
\DashArrowLine(-28.28,51.72)(-53.02,26.98){4}
\DashCArc(0,80)(60,280,310){2}
\put(-3,168){$\rho_0$}
\put(84,79){$\rho_0$}
\put(-93,79){$\rho_0$}
\put(60,139){$\rho_0$}
\put(-69,139){$\rho_0$}
\put(-67,19){$\rho_0$}
\end{picture}
\begin{picture}(100,130)(0,-30)
\Photon(0,80)(40,80){4}{5}
\Photon(1.4,80)(41.4,80){4}{5}
\put(49,77){$\equiv$}
\Photon(65,80)(105,80){4}{5}
\put(111,77){$+$}
\Photon(125,80)(185,80){4}{7}
\DashArrowLine(155,80)(155,115){4}
\put(150,124){$\rho_1$}
\put(60,8){$+$}
\Photon(75,10)(165,10){4}{11}
\DashArrowLine(103,10)(103,45){4}
\DashArrowLine(136,10)(136,45){4}
\put(98,54){$\rho_1$}
\put(131,54){$\rho_1$}
\put(170,8){$+\; \cdots$}
\end{picture}
\end{center}
\vspace{-10pt}
\caption{\em We sum diagrams with an arbitrary number of $\rho_0$ insertions using a dressed propagator 
that includes $\rho_1$ insertions.}
\label{fig:sum}
\end{figure}
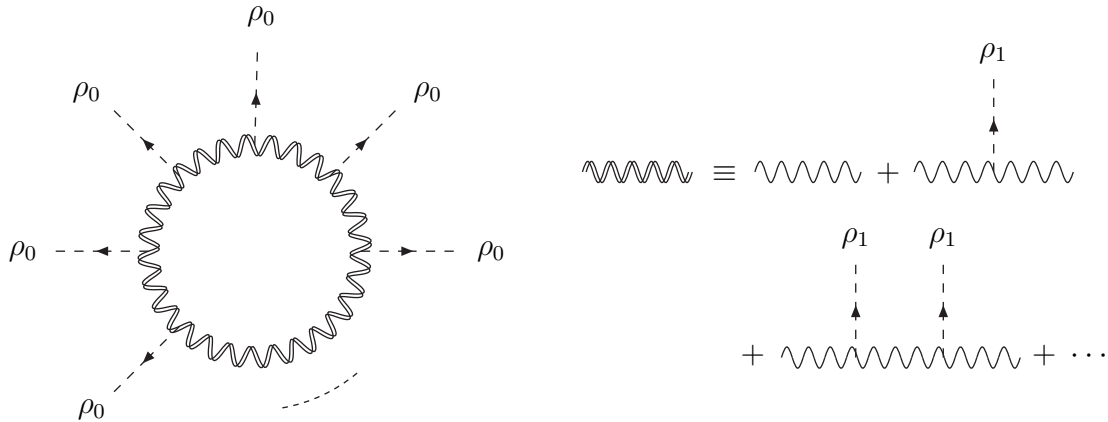

Now that we have the expression for the propagators in the absence of localized kinetic terms ($\rho_i=0$), we can 
compute the full effective  K\"ahler potential at $1$-loop, including the $\rho_i$'s. To do so, we continue to Euclidean 
momentum and follow ref.~\cite{Rattazzi:2003rj}. 
In summary, we first calculate the sum of diagrams with an arbitrary number of insertions of one of the two localized
kinetic terms, say $\rho_0$, and the other one turned off. The result depends on the brane-to-brane propagator $\Delta_{00}(p)$. 
We then replace this propagator with a propagator dressed with an arbitrary number of insertions of the other kinetic term, 
that is $\rho_{1}$ (see fig \ref{fig:sum}). Finally, we also add the diagram with no insertion of localized 
kinetic terms at all, which is proportional to $\ln Z(p)$.
Since we have already included a factor $(k z_i)^{-\frac 32}(k z_j)^{-\frac 32}$ in the definition of $\Delta_{ij}$ 
compared to the standard propagator defined with a pure $\delta$-function source and no induced metric factor, 
and the interaction localized at $z_i$ in (\ref{Lagphi}) involves a factor $(k z_i)^{-2}$, each factor $\rho_i$ will 
come along with a factor $k z_i$. The result can be written in the following very simple form as a two by two determinant:
\bea
\label{oneloopdiv}
\Omega_{\rm 1-loop} \a=\a \int\! \frac{d^4p}{(2\pi)^4} \frac{-2}{p^2} \ln \! \bigg\{\!Z(p) \det \!\bigg[
\mathbf{1}- kp^2 \!\bigg(\! \begin{array}{cc}\rho_0 z_0 \Delta_{00}(p) \!&\! \rho_0 z_0 \Delta_{01}(p) \smallskip \cr
\rho_1z_1 \Delta_{10}(p) \!&\! \rho_1z_1 \Delta_{11}(p)\end{array}\!\!\bigg)\!\bigg]\!\bigg\} \;.\qquad
\label{eq:effaction}
\eea
The momentum integral is divergent, and as expected the divergence corresponds to a renormalization of local operators.
In order to disentangle the finite corrections associated to non-local quantities that we want to compute 
from the divergent contributions corresponding to local terms, we must first classify the latter and then 
properly subtract them. 

The expected general form of UV divergences can be deduced by inspecting eqs.~(\ref{effectiveradionaction}) 
and (\ref{effectivematteraction}), which define the general form of the tree-level effective action and therefore of the 
allowed local terms. Defining for convenience $t_{0,1} = 1/z_{0,1}^2$, which as mentioned above should be thought of 
in terms of the corresponding superfield compensators, this is given by
\be
\Omega_{\rm UV} = F_0(\Phi_0)\, t_0 + F_1(\Phi_1) \, t_1 \;.
\label{UVdivs}
\ee
These divergent terms can come from two different sources: the renormalization of the 5D Planck mass
and the renormalization of the kinetic functions at the boundaries.  On the other hand, covariance under 
the Weyl shift $t_{0,1} \to \lambda^{-1/2} t_{0,1}$ discussed above constrains the whole K\"ahler function to 
have the general form
\be
\Omega = t_0\,\omega(t_1/t_0) \;.
\ee
We have not displayed the dependence on the boundary matter fields, as that is not constrained by Weyl 
symmetry. By the structure of the UV divergences in eq.~(\ref{UVdivs}), it follows that the derivative quantity
\be
t_0 \partial_{t_0} \partial_{t_1} \Omega = - \frac{t_1}{t_0} \omega^{\prime\prime}(t_1/t_0)
\ee
must be finite, since its annihilates eq.~(\ref{UVdivs}). So one way to proceed is to first calculate 
$\omega^{\prime\prime}$ and then reconstruct the full $\omega$ by solving an ordinary second order 
differential equation. This solution is determined up to two integration constants associated to the general 
solution of the homogeneous equation $\omega''=0$: $ \omega= F_0 + F_1\,t_1/t_0$. These constants 
precisely parametrize, as they should, the UV divergences in eq.~(\ref{UVdivs}).

In what follows, however, we will not directly apply the above derivative method. We shall instead subtract
from the loop integral which defines $\Omega_{\rm 1-loop}$ a suitable divergent integral with the properties 
that: 1) its functional dependence on $t_{0,1}$ is such that it is manifestly annihilated by the operator 
$\partial_{t_0}\partial_{t_1}$, 2) its subtraction makes the integral converge. The finite result obtained in 
this way is then guaranteed to contain all the finite calculable terms we are interested in. As we will explain
in more details below, the functions defining the appropriate subtractions are obtained by simply replacing 
each propagator $\Delta_{ij}(p)$ with its asymptotic behaviors $\tilde \Delta_{ij}(p)$ for $p\to\infty$. Up to 
exponentially suppressed terms of order $e^{-2p(z_1-z_0)}$, which are clearly irrelevant, we find:
\bea
\tilde\Delta_{00}(p) \a=\a \frac {1}{2p} \frac {\hat K_2(p z_0)}{\hat K_1(p z_0)}
\simeq \frac{1}{2p}\left (1+\frac{3}{2pz_0}+\frac{3}{8(pz_0)^2} + \dots\right) \;, \\
\tilde \Delta_{11}(p) \a=\a \frac {1}{2p} \frac {\hat I_2(p z_1)}{\hat I_1(p z_1)}
\simeq \frac{1}{2p}\left (1-\frac{3}{2pz_1}+\frac{3}{8(pz_1)^2} +\dots\right) \;,\\[3mm]
\tilde \Delta_{01,10}(p) \a=\a 0 \;.
\eea
Similarly, for the quantity $Z(p)$, the appropriate subtraction that has to be done to isolate the finite contribution 
associated to non-local quantities is defined by the asymptotic behavior $\tilde Z(p)$ of this quantity for $p\to \infty$. 
In this limit, the second term in \eqref{eq:Z} is of order $e^{-2p(z_1-z_0)}$ with respect  to the first, and can be neglect. Therefore, 
the quantity that controls the UV divergences in the Casimir energy is
\bea
\label{tildez}
\tilde Z(p) \a=\a \hat I_1(p z_1) \hat K_1(p z_0) \nn \\ 
\a\simeq\a \frac 12 e^{p(z_1-z_0)} \Big[1 - \frac 3{8} \Big(\frac 1{pz_1} - \frac 1{pz_0} \Big) 
 - \frac {15}{128} \Big(\frac 1{(pz_1)^2} - \frac 1{(pz_0)^2} \Big) + \cdots \Big]\,.\;\;
\eea
Notice that the contribution to the effective action related to $Z$ is proportional to an integral of $\ln Z$. 
The divergent subtraction defined by (\ref{tildez}) therefore splits as expected into the sum of two terms 
depending only on $z_0$ and $z_1$.

The non-local corrections to the K\"ahler potential can now be computed by subtracting from 
eq.~\eqref{oneloopdiv} the same expression but with $Z$ and $\Delta_{ij}$ replaced 
by their asymptotic behaviors $\tilde Z$ and $\tilde \Delta_{ij}$, that is the quantity 
\begin{equation}
\label{eq:substraction}
\Omega_{\text{div}}=  \int\! \frac{d^4 p}{\left(2 \pi \right)^4} \frac{-2} {p^2} 
\bigg[\ln \tilde Z(p) + \sum_i \ln \left(1-k p^2 \rho_i z_i \tilde \Delta_{ii}(p) \right)\bigg] \;.
\end{equation}
This formal expression, being the sum of terms that depend either on $z_0$ or on $z_1$ but not on both, 
vanishes under the action of $\partial_{t_0}\partial_{t_1}$, and is therefore an acceptable subtraction. 
Our final result is then given by
\bea
\Delta \Omega \a=\a \int  \frac {d^4p}{(2 \pi)^4} \frac {-2}{p^2}\, 
{\rm ln}\, \frac {Z(p)}{\tilde Z(p)} 
\frac {\prod_i\Big(1 - k z_i \rho_i p^2 \Delta_{ii}(p)\Big) 
- \prod_i \Big(k z_i \rho_i p^2 \Delta_{ii^\prime}(p)\Big)}
{\prod_i\Big(1 - k z_i \rho_i p^2 \tilde \Delta_{ii}(p)\Big)} \;,\qquad
\label{result}
\eea
where, as already said:
\be
\rho_i = \frac {\Phi_i^\dagger \Phi_i}{3 M_5^3} - \frac {M_i^2}{M_5^3} \;.
\ee
This formula (\ref{result}) is the main result of this paper. It generalizes the flat case result (6.32) of 
ref.~\cite{Rattazzi:2003rj} to the warped case and correctly reduces to the latter in the limit 
$k\ll 1/R$.\footnote{Indeed, in this limit the propagators $\Delta_{ij}$ and the spectral function $Z$ 
correctly reproduce the flat expressions:
\be
\lim_{kR\to 0} \Delta_{00,11} = \frac 1{2p} \coth (\pi p R) \;,\;\;
\lim_{kR\to 0} \Delta_{01,10} = \frac 1{2p} \csch (\pi p R) \;,\;\;
\lim_{kR\to 0} Z = \sinh (\pi p R) \;.
\ee
Similarly, the tilded quantities involved in the subtraction correctly reduce to the large volume
limit of the above untilded expressions.} Notice also that it is finite for any $k$, since untilded and 
tilded quantities differ by exponentially suppressed terms at large momentum.

\subsection{Structure of divergences}
It is worth spending a few words on the structure of the UV divergences we have subtracted.
Since (\ref{eq:substraction}) satisfies $\partial_{t_0}\partial_{t_1}\Omega_{\text{div}}=0$, it must 
have the same form than (\ref{UVdivs}), if properly (i.e. covariantly) regulated. It is instructive to 
see how this dependence comes about when working with a hard momentum cut-off. Let us 
first concentrate on the term proportional to $\ln \tilde Z$. Since $\tilde Z(p)$ is the product 
of a function of $pz_0$ times a function of $pz_1$, one can split the subtraction in two pieces 
depending only on $z_0$ and $z_1$, and change integration variables respectively to $v_0=p z_0$ 
and $v_1=p z_1$ in the two distinct contributions: 
\begin{equation}
\label{uvint}
\Omega_{\rm div}^{\rm sugra} =
\frac{1}{z_0^2}\int\! \frac{d^4 v_0}{\left(2 \pi \right)^4} \frac{-2}{v_0^2}  \ln \hat K_1(v_0) 
+ \frac{1}{z_1^2}\int\! \frac{d^4 v_1}{\left(2 \pi \right)^4} \frac{-2}{v_1^2} \ln \hat I_1(v_1) \;.
\end{equation}
These quantities have indeed the expected structure provided the cut-off's $\Lambda_0L$
and $\Lambda_1L$ of the $v_0$ and $v_1$ integrals do not depend on $z_0$ and $z_1$. This has 
an obvious interpretation, related to the fact that the above two distinct contributions must be 
associated (after an integral over the fifth dimension) with divergences at the two distinct 
boundaries. The point is that the original momentum integration variable $p$ is not the physical 
coordinate-invariant momentum. At each point in the bulk, the physical momentum is rather 
given by $p_{\rm phys}=\sqrt{p_\mu p_\nu g^{\mu\nu}}=pz/L$, so that the variables $v_0$ and $v_1$ 
do indeed parametrize the appropriate physical virtual momentum at each boundary. Since a 
covariant cut-off procedure should bound the physical rather than the comoving momentum,
it is indeed the cut-off for $v_{0,1}$ rather than for $p$ that must be fixed and universal. It is easy 
to check that this is indeed what happens when the above integrals are regulated through the 
introduction of 5D Pauli--Villars fields of mass $\Lambda$: one finds 
$\Lambda_0=\Lambda_1=\Lambda$. By using the asymptotic expansion of
$\hat K_1$ and $\hat I_1$ at large argument, we finally find that the divergent part 
has the form
\be
\Omega_{\rm div}^{\rm sugra} =
-3 \left (\frac{1}{z_0^2}-\frac{1}{z_1^2}\right ) \Big[\alpha (\Lambda L)^3 + \beta \Lambda L \Big]
-3 \left (\frac{1}{z_0^2}+\frac{1}{z_1^2}\right ) \gamma \ln \Lambda \;.
\ee
This expression is manifestly invariant, as it should, under the exchange of the two boundaries: 
$z_0\to z_1$, $L\to -L$. Comparing with eq.~(\ref{kahlereff}) we see that the first part corresponds 
to a renormalization $\delta M_5^3= \alpha \Lambda^3 + \beta \Lambda k^2$ of the 5D Planck mass, 
and the second part to a renormalization $\delta M_i^2 = \gamma\, k^2\ln \Lambda$ of the boundary 
kinetic terms.

Consider next the two terms in the sum of the second term of eq.~\eqref{eq:substraction}. 
Since $p \tilde \Delta_{ii}(p)$ is actually a function of $p z_i$ only and not $p z_{i^\prime}$, one 
can as before factorize all the dependence on $z_i$ from each integral be changing the 
momentum variables to $v_i = p z_i$:
\bea
\label{uvintbis}
\Omega_{\rm div}^{\rm mat} \a=\a
\frac{1}{z_0^2}\int\! \frac{d^4 v_0}{\left(2 \pi \right)^4} \frac{-2}{v_0^2}  
\ln \bigg[1 - \frac {k \rho_0 v_0}2 \frac {\hat K_2(v_0)}{\hat K_1(v_0)} \bigg] \nn \\
\a\;\a +\, \frac{1}{z_1^2}\int\! \frac{d^4 v_1}{\left(2 \pi \right)^4} \frac{-2}{v_1^2} 
\ln \bigg[1 - \frac {k \rho_1 v_1}2 \frac {\hat I_2(v_1)}{\hat I_1(v_1)} \bigg] \;.
\eea
The same arguments as before concerning covariance and the appropriate cut-off's to be used
apply. The result is again that the integral can be evaluated by using the asymptotic expression 
for large argument of the functions appearing in its integrand and the dimensionless cut-off 
$\Lambda L$ for both variables $v_i$. Treating $\rho_i$ as small quantities, we can also expand 
the logarithms. Proceeding in this way we finally find:
\bea
\label{uvintbisbis}
\Omega_{\rm div}^{\rm mat} \a=\a
\sum_{n=1}^{\infty} \frac{\rho_0^n}{z_0^2} \Big[\alpha_{n} \Lambda^{n+2} L^2 + \cdots \Big]
+ \sum_{n=1}^{\infty} \frac{\rho_1^n}{z_1^2} \Big[\alpha_{n} \Lambda^{n+2} L^2 + \cdots \Big] \;.
\eea
Comparing with eq.~(\ref{kahlereff}) we see that the $n$-th term corresponds to a correction to the 
4D potentials $\Omega_i$ of the form $\delta \Omega_i = \alpha_n \Lambda^{n+2}
(3 M_5^{3})^{-n} (-3 M_i^2 + \Phi_i \Phi_i^\dagger)^n + \cdots$, the dots representing less 
singular terms. This correction represents a renormalization of
the quantities $M_i^2$, the wave function multiplying the kinetic term 
$\Phi_i \Phi_i^\dagger$ of the matter fields and the coefficients of those
higher-dimensional local interactions involving up to $n$ powers of 
$\Phi_i \Phi_i^\dagger$.

\subsection{General Results}
Let us now come to a more explicit discussion of the results.
Comparing eq.~(\ref{result}) with the general expression (\ref{corrkahlereff}),
it is possible to extract all the coefficients $C_{n_0,n_1}$. We will consider in 
more details the first coefficients with $n_{0,1}=0,1$, which control the vacuum 
energy and the scalar soft masses that are induced by supersymmetry breaking, as 
functions of the quantities 
$$
\alpha_i = \frac {M_i^2}{M_5^3} \;,
$$ 
which define the localized kinetic terms for the bulk fields. Taking suitable derivatives 
of (\ref{result}) with respect to $\rho_i$ and setting these to $-\alpha_i$, we find:
\bea
C_{0,0} \a=\a -2 \int \frac {d^4p}{(2 \pi)^4} p^{-2}\,
\Big[{\rm ln}\,{\bf Z}(p) - {\rm ln}\,\tilde {\bf Z}(p)\Big] 
\label{c00} \;, \\ 
C_{1,0} \a=\a \frac 2{3M_5^3} (k z_0) \int \frac {d^4p}{(2 \pi)^4}\,
\Big[{\bf \Delta}_{00}(p) - \tilde {\bf \Delta}_{00}(p)\Big]
\label{c10} \;, \\
C_{0,1} \a=\a \frac 2{3M_5^3} (k z_1) \int \frac {d^4p}{(2 \pi)^4}\,
\Big[{\bf \Delta}_{11} (p) - \tilde {\bf \Delta}_{11}(p)\Big]
\label{c01} \;, \\
C_{1,1} \a=\a \frac 2{9 M_5^6} (k z_0) (k z_1) \int \frac {d^4p}{(2 \pi)^4}\,p^2\,
\Big[{\bf \Delta}_{01} (p) {\bf \Delta}_{10} (p) \Big]
\label{c11} \;,
\eea
where ${\bf \Delta}_{ij}$ are the brane-to-brane propagators in
presence of localized kinetic terms $\alpha_i$,
and ${\bf Z}$ is the $\alpha_i$-dressed analog of $Z$:
\bea
{\bf Z} \a=\a Z \Big[\Big(1 + k z_0 \alpha_0 p^2 \Delta_{00}\Big)
\Big(1 + k z_1 \alpha_1 p^2 \Delta_{11}\Big)
- k z_0 \alpha_0 k z_1 \alpha_1 p^4 \Delta_{01} \Delta_{10}\Big] \;,\qquad \\
{\bf \Delta}_{00} \a=\a \frac {\Delta_{00}
\Big(1 + k z_1 \alpha_1 p^2 \Delta_{11}\Big) 
- k z_1 \alpha_1 p^2\Delta_{01} \Delta_{10}}
{\Big(1 + k z_0 \alpha_0 p^2 \Delta_{00}\Big)
\Big(1 + k z_1 \alpha_1 p^2 \Delta_{11}\Big)
- k z_0 \alpha_0 k z_1 \alpha_1 p^4 \Delta_{01} \Delta_{10}} \;, 
\raisebox{32pt}{} \\
{\bf \Delta}_{11} \a=\a \frac {\Delta_{11}
\Big(1 + k z_0 \alpha_0 p^2 \Delta_{00}\Big) 
- k z_0 \alpha_0 p^2 \Delta_{01} \Delta_{10}}
{\Big(1 + k z_0 \alpha_0 p^2 \Delta_{00}\Big)
\Big(1 + k z_1 \alpha_1 p^2 \Delta_{11}\Big)
- k z_0 \alpha_0 k z_1 \alpha_1 p^4 \Delta_{01} \Delta_{10}} \;,\\
{\bf \Delta}_{01,10} \a=\a \frac {\Delta_{01,10}}
{\Big(1 + k z_0 \alpha_0 p^2 \Delta_{00}\Big)
\Big(1 + k z_1 \alpha_1 p^2 \Delta_{11}\Big)
- k z_0 \alpha_0 k z_1 \alpha_1 p^4 \Delta_{01} \Delta_{10}} \;.
\raisebox{22pt}{}
\eea
The subtractions $\tilde{\bf{Z}}$ and $\tilde{\bf{\Delta}}_{ij}$  are 
similarly defined out of $\tilde Z$ and $\tilde \Delta_{ij}$.

\subsection{Results in the absence of localized kinetic terms}

In the simplest case of vanishing localized kinetic terms, that is $\alpha_i = 0$,
the first four relevant terms in the correction to the effective K\"ahler potential 
are given by eqs.~(\ref{c00})--(\ref{c11}) with ${\bf Z} \rightarrow Z$ and 
${\bf \Delta}_{ij} \rightarrow \Delta_{ij}$, and similarly for the tilded quantities 
defining the subtractions. 

In the limit of flat geometry, one recovers the known results for the coefficients
of the four leading operators:
\bea
\label{cflat}
\begin{array}{ll}
\displaystyle C_{0,0} = \frac {c_{\rm f}}{4 \pi^2} \frac 1{(T + T^\dagger)^2} 
\label{c00flat} \;, &
\displaystyle C_{1,0} = \frac {c_{\rm f}}{6 \pi^2 M_5^3} \frac 1{(T + T^\dagger)^3}
\label{c10flat} \;, \cr \\[-3mm]
\displaystyle C_{0,1} = \frac {c_{\rm f}}{6 \pi^2 M_5^3} \frac 1{(T + T^\dagger)^3} 
\label{c01flat} \;, &
\displaystyle C_{1,1} = \frac {c_{\rm f}}{6 \pi^2 M_5^6} \frac 1{(T + T^\dagger)^4} 
\label{c11flat} \;,
\end{array}
\eea
where $c_{\rm f} = \zeta(3) \approx 1.202$. Let us mention that these results
differ from the previous results of ref.~\cite{Rattazzi:2003rj,Buchbinder:2003qu} 
by a factor of $2$ mismatch in the definition of the 5D kinetic coefficient $M_5^2$. Concerning ref.~\cite{Rattazzi:2003rj}, we found that the source of the discrepancy is a wrongful normalization of the graviton propagator employed in that paper.
Moreover, the argument we shall give in next section  
confirms  in a simple and unambiguous way that the correct result is the one presented just above.

In the limit of very warped geometry the coefficients of the first
four operators are instead:
\bea
\label{cwarp}
\begin{array}{ll}
\displaystyle C_{0,0} = \frac {c_{\rm w} k^2}{4 \pi^2} e^{-2k(T + T^\dagger)} 
\label{c00warp} \;, &
\displaystyle C_{1,0} = \frac {c_{\rm w} k^3}{12 \pi^2 M_5^3} e^{-2k(T + T^\dagger)} 
\label{c10warp} \;, \cr \\[-3mm]
\displaystyle C_{0,1} = \frac {c_{\rm w} k^3}{6 \pi^2 M_5^3} e^{-2k(T + T^\dagger)} 
\label{c01warp} \;, &
\displaystyle C_{1,1} = \frac {c_{\rm w} k^4}{18 \pi^2 M_5^6} e^{-2k(T + T^\dagger)}
\label{c11warp} \;,
\end{array}
\eea
where
\be
c_{\rm w} = \frac 12 \int_0^\infty dx \, x^3 \frac {K_1(x)}{I_1(x)} 
= \frac 18 \int_0^\infty  \frac {dx \, x^3}{I_1(x)^2} \approx 1.165 \;.
\ee

\subsection{Results in the presence of localized kinetic terms for the case of small and large warping}

In the presence of localized kinetic terms, that is $\alpha_i \neq 0$, it is 
convenient to consider directly the full K\"ahler potential as a function 
$\rho_i = - \alpha_i + \Phi_i \Phi_i^\dagger/(3 M_5^3)$. 

In the flat case, the result simplifies to
\be 
\Delta \Omega = \frac 1{4 \pi^2} \frac 1{(T + T^\dagger)^2} 
f_{\rm f}(\frac {\rho_0}{T + T^\dagger},\frac {\rho_1}{T + T^\dagger})
\ee
where  
\be
f_{\rm f}(a_0,a_1) = - \int_0^\infty dx\, x\, {\rm ln} 
\Bigg[1 - \frac{1 + a_0\,x/2}{1 - a_0\,x/2}\,
\frac{1 + a_1\,x/2}{1 - a_1\,x/2}\, e^{-x}\Bigg] \;.
\label{fflat}
\ee
Expanding this expression at leading order in the matter fields, one 
finds\footnote{We use the standard notation 
$f^{(n_0,n_1)}(a_0,a_1) = (\partial/\partial a_0)^{n_0}(\partial/\partial a_1)^{n_1}
f(a_0,a_1)$.} 
\be
C_{n_0,n_1} = \frac {1}{4\cdot 3^{n_0+n_1} \pi^2 M_5^{3 n_0 + 3 n_1}}
\frac 1{(T + T^\dagger)^{2+n_0+n_1}} f_{\rm f}^{(n_0,n_1)}\Big(\frac {-\alpha_0}{T + T^\dagger},
\frac {-\alpha_1}{T + T^\dagger}\Big)\;.
\ee
As already discussed in ref.~\cite{Rattazzi:2003rj}, $C_{0,0}$ and $C_{0,1}$ become negative for 
large $\alpha_0$ and small $\alpha_1$. Similarly $C_{0,0}$ and $C_{1,0}$ become negative 
for large $\alpha_1$ and small $\alpha_0$.

In the limit of large warping, one finds:
\be
\Delta \Omega = \frac {k^2}{4 \pi^2} e^{-2k(T + T^\dagger)} f_{\rm w}(k \rho_0,k \rho_1)
\ee
where  now
\be
f_{\rm w}(a_0,a_1) = \frac 12 \int_0^\infty dx\, x^3\, 
\frac {K_1(x)}{I_1(x)} \frac 1{1 - a_0} 
\frac {1 + a_1\,x/2\, K_2(x)/K_1(x)}{1 - a_1\, x/2\, I_2(x)/I_1(x)} \;.
\label{fwarp}
\ee
From this we deduce that
\be
C_{n_0,n_1} = \frac {k^{2+n_0+n_1}}{4\cdot 3^{n_0+n_1} \pi^2 M_5^{3 n_0 + 3 n_1}}
e^{- 2 k (T + T^\dagger)} f_{\rm w}^{(n_0,n_1)}(-\alpha_0 k,-\alpha_1 k)\;.
\label{cwarped}
\ee
None of the coefficients becomes negative for large $\alpha_0$ and small $\alpha_1$, whereas 
$C_{0,0}$ and $C_{1,0}$ become negative for $\alpha_1 \sim 0.6/k$ and small $\alpha_0$. We again 
stress that there is an upper bound for $\alpha_1$, as for $\alpha_1 =1/k$, the radion becomes a ghost. 
But luckily, $C_{1,0}$ changes sign before $\alpha_1$ reaches this critical value.

Notice that the dimensionless parameters that control the effect of the localized kinetic terms depends
on the warping. In the flat limit, they are given by $\alpha_i/(2 \pi R)$, and their impact therefore significantly 
depends on the radius dynamics, whereas in the large warping limit they are given by $\alpha_i k$, and their
impact is therefore not directly dependent of the radius. 

\section{A simplified computation}\setcounter{equation}{0}

We present here an alternative technique which allows to rederive in an immediate way
some of the results of the present paper and of refs.~\cite{Gherghetta:2001sa,Rattazzi:2003rj,Buchbinder:2003qu}.
The basic remark is that the effects we are computing are dominated by relatively soft modes with 5D
momentum of the order of the compactification scale. The softness of these modes does not allow them to 
distinguish between an infinitesimal shift in the position of a boundary, i.e.\ a small change of the radius, 
and the addition of infinitesimal boundary kinetic terms. The equivalence between these two effects\footnote{This 
equivalence is related to the analysis of ref.~\cite{Lewandowski:2002rf}, which shows that a displacement of the
branes can be accounted for by a renormalization group flow of the local operators that they support.} allows one
to relate in a straightforward way the matter-independent term (Casimir energy) to the matter-dependent ones
(radion and brane-to-brane mediation terms).

Let us see how this works in more detail. As it will become clear from the discussion, warping and spin play no role. 
Therefore we can focus first on the case of the scalar Lagrangian in eq.~(\ref{Lagphi}) in the flat limit $k\to 0$. 
Notice that $\rho_{0,1}$ have the dimension of a length, so that the case of infinitesimal kinetic terms should 
correspond to $\rho_{0,1}\ll \pi R$. Consider the equations of motion and boundary conditions in the vicinity of 
one of the two boundaries, say $y=0$:
\begin{equation}\label{eq:KK} 
\varphi''(y) = - p^2 \varphi(y)\;,\;\;\;\;
\varphi'(0) = \frac{\rho_0}{2} p^2 \varphi(0) \;.
\end{equation}
Now, for $p \rho_0\ll 1$ the solution does not vary appreciably over the 
distance $\rho_0$. Therefore we can safely write the value of $\varphi'$ at the point $y=\rho_0/2$ by means 
of a Taylor expansion. By using both eqs.~(\ref{eq:KK})  we obtain then
\bea
\varphi'(\frac{\rho_0}{2})\a=\a\varphi'(0)+\varphi''(0)\frac{\rho_0}{2}+\varphi'''(0)\frac{\rho_0^2}{8}+\cdots \nn \\
\a=\a 0+\mathcal{O}(\rho_0^2 p^2)\varphi'(0) +\mathcal{O}(\rho_0^3 p^4)\varphi(0)+\cdots \;.
\eea
This equation implies that up to terms of order $\rho_0^2 p^2$, the solutions are the same 
as those of an equivalent system with Neumann boundary condition $\varphi'=0$ at a shifted boundary $y'_0=\rho_0/2$.
Consequently, at the same order in $\rho_0$ all the properties of the two systems, KK masses included, will be the same.
The same remark applies, in a totally independent way, to the other boundary.  Now, quantities like the Casimir energy are dominated by
modes of momentum $p\sim 1/R$, so that the above equivalence works up to terms of order $\rho_0^2/R^2$, $\rho_1^2/R^2$.

This means that radion mediation $\mathcal{O}(\rho_0)$ and brane-to-brane mediation $\mathcal{O}(\rho_0\rho_1)$ effects can be 
read off by simply shifting the radius as $2\pi R\to 2\pi R-\rho_0-\rho_1$ in the Casimir energy (or the K\"ahler function). In fact the 
shifted quantity $M^2=M_5^3( 2\pi R-\rho_0-\rho_1)$ is nothing but the effective low energy Planck mass   
in the presence of boundary kinetic terms. Being a low energy quantity, $M^2$ manifestly  
realizes the equivalence between boundary kinetic terms and shift in $R$. 

The argument we used above is very robust and quickly generalizes to other cases of interest. In particular,
as the argument is based on a local expansion close to the boundary where the small kinetic term has been added, 
it still holds true whatever boundary condition is given at the other boundary. For instance, in the
presence of a sizable kinetic term $\rho_1$, turning on an infinitesimal $\rho_0$ is still equivalent to the shift
$2\pi R\to 2\pi R-\rho_0$ in the full Casimir energy $V(R,\rho_1)$. Moreover the presence of curvature is
clearly not affecting our basic argument as long as $\rho_{0,1}$ are smaller than the typical curvature length
of the metric. Basically, curvature introduces an extra, but fixed, momentum scale in addition to the 4D momentum $p$
of our previous argument. For the RS metric, we would have a double expansion in both $\rho_0 p$ and $\rho_0 k$.
Without doing an explicit computation it is easy to deduce the equivalence relation between boundary kinetic terms and
brane shift in the RS model. The point is that the effective 4D Planck mass
\be
M^2 = {M_5^3}{k} \Big[\frac {1 - k \rho_0}{z_0^2} 
- \frac{1 + k  \rho_1}{z_1^2}\Big] 
\ee
must be invariant. We conclude that, at linear order, switching on infinitesimal 
$\rho_i$'s is equivalent to shifting the positions $z_i$ of the latter to
\bea
\label{shifts}
z_0^\prime \a=\a z_0 e^{+ k  \rho_0/2} \simeq z_0 + k  \rho_0 z_0/2 \;,\nn \\
z_1^\prime \a=\a z_1 e^{- k  \rho_1/2} \simeq z_1 - k  \rho_1 z_1/2 \;.
\eea

\subsection{Flat case}

Let us consider now in more detail the implications of the above reasoning. 
In the flat case, without localized kinetic terms, the matter-independent 
effective K\"ahler potential is very simple to deduce (see also ref.~\cite{Ponton:2001hq}). The modes
have Neumann boundary conditions at both branes and masses $m_n = n/R$. One then finds
$\Delta\Omega \simeq c_{\rm f}/(4\pi^2)\,(T+T^\dagger)^{-2}$, where $c_{\rm f}= {\rm Li}_3(1) = \zeta(3)$. 
Applying to this result the shift $T + T^\dagger \to T + T^\dagger 
- (\Phi_0^\dagger \Phi_0 + \Phi_1^\dagger \Phi_1)/(3 M_5^3)$ we find:
\begin{equation}\label{eq:old}
\Delta\Omega \simeq \frac{\zeta(3)}{4\pi^2} \Big[T+T^\dagger - \frac {\Phi_0^\dagger \Phi_0}{3 M_5^3} 
- \frac {\Phi_1^\dagger \Phi_1}{3 M_5^3}\Big]^{-2} \;.
\end{equation}
Expanding this expression to linear order in each brane term, we reproduce the coefficients (\ref{cflat})
for the four leading operators.\footnote{We thank Adam Falkowski for pointing out this fact to us thus stimulating 
the discussion of this section.}

It is also illuminating to consider the limit in which one of the kinetic terms, say $\rho_1$, becomes large
with the visible sector located at $y=0$. The condition at $y=\pi R$ reduces to a Dirichlet boundary condition, forcing 
all the wave functions of the non-zero modes to vanish there. The massive spectrum
becomes $m_n = (n+1/2)/R$, but with the zero mode mass obviously unaltered. The matter-independent term in the K\"ahler
potential is then given by $\Delta\Omega \simeq c_{\rm f}/(4\pi^2)\,(T+T^\dagger)^{-2}$, where now $c_{\rm f}= 
{\rm Re}\,{\rm Li}_3(e^{i \pi}) = -3/4 \zeta(3)$. Knowing this result, and
applying again the shift, which now implies $T + T^\dagger \to T + T^\dagger 
- \Phi_{0}^\dagger \Phi_{0}/(3 M_5^3)$, we can deduce that:
\begin{equation}
\label{eq:oldlarge}
\Delta\Omega \simeq - \frac 34
\frac{\zeta(3)}{4\pi^2} \Big[T+T^\dagger - \frac {\Phi_{0}^\dagger\Phi_{0}}{3 M_5^3} \Big]^{-2} \;.
\end{equation}
When expanded, this again agrees with our result for the two leading operators. As remarked in ref.~\cite{Rattazzi:2003rj}, the 
effect of the large localized kinetic term is to flip the sign of the coefficient of the operator controlling radion mediation, and 
to send the coefficient of the operator controlling the brane-to-brane mediation to zero. 

Notice finally that ref.~\cite{Rattazzi:2003rj} 
also considered the case in which the radion $T$ is stabilized precisely by the 1-loop Casimir energy in the presence of localized kinetic terms. By placing the visible 
sector on a brane where no other kinetic term contribution was present, it was then found, surprisingly, that the soft masses 
vanished exactly at the minimum of the radion potential. Our argument about the shift symmetry makes this result obvious.
The scalar mass term is obtained by shifting $T + T^\dagger \to T + T^\dagger - \Phi_{0}^\dagger \Phi_{0}/(3 M_5^3)$ in the 
radion Casimir energy $V(T+T^\dagger)$. The scalar mass is therefore proportional to $V'$, and vanishes at the minimum of 
the potential.

\subsection{Strongly warped case}

To conclude we can consider the strongly warped case. In the absence of localized kinetic terms, the matter-independent 
term in the effective K\"ahler potential is known from refs.~\cite{Garriga:2000jb,Garriga:2002vf}  (see also \cite{Saharian:2002bw,Goldberger:2000dv,Brevik:2000vt} )and has the form 
$\Delta\Omega_{\rm eff} \simeq c_{\rm w}/(4\pi^2)\,z_0^2 z_1^{-4}$, where $c_{\rm w} \simeq 1.165$. Applying the shift, 
we find 
\be
\Delta \Omega \simeq \frac {c_{\rm w}\,k^2}{4 \pi^2}
\exp \Bigg\{-2k\Bigg[T + T^\dagger - \frac 16 \frac {\Phi_0^\dagger \Phi_0 }{M_5^3}
- \frac 13 \frac {\Phi_1^\dagger \Phi_1 }{M_5^3} \Bigg]\Bigg\} \;.
\ee
which, when expanded at linear order in $\Phi_0 \Phi_0^\dagger$ and in $\Phi_1 \Phi_1^\dagger$, reproduces 
our results (\ref{cwarp}) for  $C_{1,0}$, $C_{0,1}$ and $C_{1,1}$.

In this case, no illuminating argument is available for the effect of a localized kinetic term. More precisely, in the not very interesting 
situation in which a kinetic term is located at the UV brane, its only effect is basically to increase the effective 4D Planck mass, 
and therefore all the radiative effects get simply suppressed when it becomes large, without any sign flip. In the more interesting
situation in which a kinetic term is switched on at the IR brane, the impact on the radiative effects can instead be more interesting 
and significant, as we saw, but due to the fact that the latter is limited to stay below a finite critical value, there is no useful limit in 
which the problem simplifies. One has then to rely on the exact computation to understand its physical consequences.

\section{Applications}\setcounter{equation}{0}

We would now like to investigate the extent to which our result for the radion-mediated contribution to the scalar soft masses 
squared can help to cure the problem of tachyonic sleptons of anomaly mediation. In order to do that, we need to find a model in 
which the radius is stabilized, and supersymmetry is broken in such a way that radion mediation dominates over anomaly and 
brane-to-brane mediation. In this section, we study the effective description of such a model

By effective description, we mean that we do not fully specify the sector that breaks supersymmetry, but parametrize it through a 
Goldstone supermultiplet $X$ with a linear superpotential. Similarly,  the radion is stabilized by some 
unspecified 5D dynamics (see \cite{Luty:2000ec,Arkani-Hamed:1999pv,Luty:1999cz,Goldberger:1999un,Goh:2003yr,Maru:2003mq} for specific examples) that we shall parametrize through an effective superpotential depending on the radion multiplet.
Finally, in order to cancel the cosmological constant, we also need to add a constant superpotential and tune its 
coefficient. All these superpotentials admit microscopic realizations, for instance in terms of gaugino condensations, 
but we will not discuss them here in any detail. What is instead important for us is that in such a general situation, 
there are three important sources of supersymmetry breaking effects for the visible sector matter and gauge fields, coming from 
the $F$ terms of the compensator, the radion and the Goldstone chiral multiplets, respectively $\phi$, $T$ and $X$. 
These will induce contributions to the soft masses corresponding to anomaly, radion, and brane-to-brane mediation effects.
 The absence of tachyons requires, necessarily, that the radion contribution be positive and comparable
or bigger than the other two. We shall discuss some examples in this section. On the other hand, the extent to which a model
helps solving the supersymmetric flavor problem depends on the separation between the compactification scale
and whatever other fundamental UV scale there is in the theory, for instance  $M_5$. The bigger this separation is,
the more suppressed are higher order effects and in principle flavor-breaking corrections to soft masses. We will not give 
a full discussion of 
this problem in this paper as it will require a careful model by model analysis. We plan to come back on this issue
in a forthcoming paper \cite{grs}.

\subsection{Flat case}

In ref.~\cite{Rattazzi:2003rj}, a model of the type discussed above was considered, along the lines of \cite{Luty:1999cz}. 
The radion was stabilized 
by gaugino condensation in the bulk and on the hidden brane, giving rise to the following low energy K\"ahler potential and superpotential:
\begin{eqnarray}
\Omega \a=\a \Big[- 3 M_5^3 (T + T^\dagger)  - 3 M_1^2+ X^\dagger XZ(XX^\dagger/\Lambda_c^2)\Big] \phi^\dagger \phi \;, \\
\label{eq:flatsuper}
W \a=\a \Big[\Lambda_a^3 e^{-n \Lambda_a T} + \Lambda_b^3 + \Lambda_c^2 X \Big] \phi^3\;.
\end{eqnarray}
We are assuming for simplicity that the wave function $Z$ stabilizes $X$ at the origin $X=0$
(see for example ref.~\cite{Izawa:1997gs}).
We are also assuming that the hidden sector and the brane kinetic term $M_1^2$ are localized on the same brane.
Minimizing the potential, it is found that the radius is stabilized, and we can tune $\Lambda_c$ to cancel the cosmological constant. 
The order of magnitude of the different $F$-terms are found to be:
\begin{equation}
F_\phi \sim \frac{F_X}{M} \sim \frac{F_T}{T } (n \Lambda_a T) \;.
\end{equation}
The point here is that the radion-mediated contribution, the only one that can give positive mass squared, is suppressed
compared to the two other contributions by a factor of $(n \Lambda_a T)^{-1}$, which is precisely the loop factor 
$\alpha_G\equiv g_5^2/(16\pi^2 T)$ for the 5D gauge interactions at the compactification scale. Because of this 
extra suppression the ``equality'' between radion-mediated and anomaly-mediated contributions is obtained for
\be
\alpha_G^2\,\alpha_5 \sim \left (\frac{g}{4\pi}\right )^4\sim \alpha^2 \;,
\label{equality}
\ee
where $\alpha_5=1/(M_5 T)^3$ measures the loop expansion parameter for 5D gravity. This relation implies that the 
radius should be fairly small. Moreover, a second, necessary, requirement is that the radion contribution be positive 
and dominant with respect to the  brane-to-brane one. Adding a localized kinetic term on the hidden brane helps in 
two ways. It makes the radion-mediated contribution positive while suppressing  brane-to-brane effects. Taking into 
account the extra $\mathcal{O}(\alpha_G^2)$ suppression of the radion contribution one finds by inspection of the 
explicit formulae that a positive mass is obtained for a pretty large brane kinetic term: $M_1^2/M_5^3 T \gsim 1/\alpha_G^2$.
In the regime where the bulk gauge theory is perturbative at the compactification scale, the size of the 4D effective
Planck scale is fully determined by the boundary, with the bulk playing the just role of a small perturbation. Although 
peculiar this is not the real problem of this model. The real problem is represented by eq.~\eqref{equality}, which 
implies that the gauge and gravity expansion parameters are not so small. These parameters control higher order effects.
For instance, it is natural to expect flavor violating four-derivative boundary gravitational 
interactions, suppressed by two extra powers of $M_5$. These would give subleading 
$\mathcal{O}(1/(M_5 T)^2) \sim \alpha_5^{2/3}$ flavor-breaking corrections to the 
radion-mediated effect \cite{Rattazzi:2003rj}. Similarly, one expects extra interactions 
between the boundary matter fields and the bulk gauge fields. At the visible brane, we expect 
flavor violating couplings of the type $Q_iQ_j^\dagger W_\alpha W^\alpha$, and similarly at the 
hidden brane we expect a coupling of the fom $X X^\dagger W_\alpha W^\alpha$, suppressed 
by three powers of some high energy scale. In presence of these interactions, one-loop exchange 
of the bulk gauge fields would add to the K\"ahler potential a flavor-violating brane-to-brane term. 
If the relevant scale suppressing the couplings were the five-dimensional quantum gravity scale, 
then these effects would even be larger than the universal radion-mediated term. The presence of 
localized gauge kinetic terms for the gauge fields would help making this contribution smaller than 
the universal one, but only at the price of making the bulk gauge theory more strongly coupled.
This fact makes this pathway to model building not very appealing, though it may be worth a more 
detailed analysis. Notice however that the difficulties arise from the extra suppression of the radion 
$F$-term in the specific model we are considering. 
This suppression follows directly from the no-scale structure of the K\"ahler potential (i.e., linearity in $T$) 
but depends also crucially on the specific mechanism of radion stabilization (the superpotential). Indeed, 
assuming a general $W(T)$ and the same K\"ahler potential, the stationarity conditions are easily seen to imply
\be
\frac{F_T}{T}= F_\phi \frac{\partial_T W}{T \partial^2_T W} \;.
\label{generalft}
\ee
Now, for a generic superpotential depending on powers of $T$, we would expect the right hand side to be 
of order $F_\phi$. Indeed, this is what one finds for instance for $W=a+bT^{-n}$. However, for the purely 
exponential dependence of eq. \eqref{eq:flatsuper} there comes an extra $1/T$ suppression in $F_T/T$ 
with respect to the naively expected result. Notice that for $F_T/T\sim F_\phi$, $\alpha_G$ would drop 
out from eq.~\eqref{equality} and the expected flavor violation from gravity loops would scale like 
$\alpha_5^{2/3}\sim \alpha^{4/3}$, which is in the interesting range. It would be interesting to look for 
alternative models of radius stabilization where $F_T/T \sim F_\phi$ and we leave it for future work. 
In the meantime, we would like to study some basic features of the warped case, which, as we shall see, 
opens a perhaps more promising direction of investigation.

\subsection{Strongly warped case}

In the warped case, the situation is similar, but the two possible choices for the locations of the visible and 
hidden sectors are no longer equivalent, and must be studied separately. Notice also that the radion, 
as parametrized by $\mu = k \phi e^{-kT}$, looks like a ordinary matter field with a canonical K\"ahler potential. 
This is not at all surprising in view of the holographic interpretation. Moreover, unlike in a flat geometry, a 
constant superpotential localized on the IR brane leads to a radion-dependent effective 4D superpotential.

We will consider the case where the visible sector is on the UV brane at $z_0$ and the hidden sector on the 
IR brane at $z_1$, as in this case, localized kinetic terms can make the soft masses squared positive. If the visible 
sector is put on the IR brane, it is not possible to obtain positive corrections to the soft masses squared from the 
gravitational sector; therefore we do not study this case further. 

Since the matter and gauge fields live on the UV brane, they have a canonically normalized kinetic 
term and they couple to the ordinary conformal compensator $\phi$. The soft masses are then given by
\begin{eqnarray}
m_{1/2} \a=\a a \frac{g^2}{(4 \pi)^2} \left|F_\phi \right| 
\label{m12bis} \;, \\
m_{0}^2  \a=\a b\,\frac{g^4}{(4 \pi)^4} \left|F_\phi \right|^2\!
- \frac{\left|\mu \right|^2}{3 \pi^2 M_5^3 k} f_w^{(1,0)}\!\left(-\alpha_i k \right)  |F_\mu|^2\! 
- \frac{ k^2\left| \mu \right|^2}{ 36 \pi^2 M_5^6} f_w^{(1,1)}\! \left(-\alpha_i k \right) |F_X|^2 \;.\qquad\;\;
\label{m02bis} 
\end{eqnarray}
Notice that we canonically normalized the Goldstone multiplet $X$, see eq.~\eqref{68}.
The numerical coefficients $a$ and $b$ controlling the anomaly-mediated contributions depend on the quantum numbers 
of the corresponding particles. The function $f_w$ was calculated in section 4 (c.f. eq.~\eqref{cwarped}) in the limit of 
large warping and is a function of the localized kinetic terms. As in the flat case, $f^{(1,0)}$ and $f^{(1,1)}$ are both positive
for $\alpha_0=0$ and $\alpha_1=0$, but for $\alpha_0=0$ and $\alpha_1 k > 0.6$, the first becomes negative, whereas the 
second remains positive. We therefore have the same potentially interesting case as in the flat case for $\alpha_0$ and 
$\alpha_1 k \sim 1$, although the brane-to-brane-mediated contribution, which stays negative, is a priori not suppressed 
with respect to the radion-mediatied  contribution, which can become positive.

We parametrize the effective K\"ahler potential and superpotential as follows:
\begin{eqnarray}
\Omega \a=\a - 3  \frac{M_5^3}{k^3}  \left(k^2 \phi^\dagger \phi - \mu^\dagger\mu\right) 
- 3 M_0^2 \phi^\dagger\phi - 3 \frac{M_1^2}{k^2} \mu^\dagger\mu + X^\dagger X  \;, \label{68}\\
W \a=\a \Lambda^{3-n}\, \phi^{3-n} \mu^n + \Lambda_{0}^3\,\phi^3 + a\, \mu^3
+ b \, X \mu^2 \;.
\end{eqnarray}
The superpotential terms can arise from both gaugino condensation \cite{Luty:2000ec} and tree level dynamics,
like in the supersymmetric version of the Goldberger-Wise mechanism \cite{Goldberger:1999un,Goh:2003yr}.
Compatibly with the AdS/CFT interpretation the first term can be interpreted as coming from a deformation
by chiral operator of conformal dimension $n$. Then unitarity requires $n>1$. In fact for the case in
which this term arises from gaugino condensation \cite{Luty:2000ec} we have $n\sim 4\pi^2 /(k g_5^2) \gg 1$.
Low energy quantities depend on the localized kinetic terms through the combinations 
$M_\phi^2 = M_5^3/k + M_0^2$ and $M_\mu^2 = (M_5^3/k - M_1^2)$. The Planck mass is given
by $M^2 = M_\phi^2 - M_\mu^2 |\mu/ k|^2$, or just $M^2 \simeq M_\phi^2$ for small $\mu$.
The extreme case where $\Lambda_{0} = 0$ with $a \neq 0$, that is with a constant superpotential 
only in the hidden sector, was analyzed in ref.~\cite{Luty:2000ec}. Taking an appropriate choice for the parameters 
$\Lambda, a$ and $b$, there exists a solution with vanishing cosmological constant and 
small $\mu /k$. 
However, we get in this case that $F_\phi \sim F_\mu/\mu \sim  F_X/M$. This implies that the contribution 
from radion mediation is parametrically smaller than the one from  brane-to-brane mediation by a factor $(\mu /k)^2$.
Therefore this case cannot work in the limit of large warping that we want to consider.
The opposite extreme case, where $\Lambda_{0} \neq 0$ and $a= 0$, that is to say just a 
constant superpotential on the Plank brane,
 can be analyzed similarly. In this case,
a solution with vanishing cosmological constant and small $\mu /k$ exists but only for  $n<1$. Thus we  
shall not consider this case further. 
A more interesting situation can be obtained by considering the more general regime in which $a$ 
is non-zero and $\Lambda_{0}$ is small enough to be negligible in the radius  stabilization dynamics
(which is then entirely controlled by the positive matter terms associated to $\mu$ and $X$)
This assumption is consistent with the possibility to tune $\Lambda_0$ to cancel the cosmological constant:
this is the same remark usually made in models where supersymmetry is dynamically broken at low energy.
In such a situation, we can find a solution with vanishing
cosmological constant and small $\mu /k$ by tuning $\Lambda_0^3$ and 
by choosing $(\Lambda/ k)^{3-n} \gg a$ for $n > 3$ or $(\Lambda /k)^{3-n} \ll a$
for $n < 3$.  Defining for convenience the parameter 
\begin{equation}
\rho =\left |\frac {b\, M_\mu }{k a}\right| \;,
\end{equation}
we find that the cancellation of the cosmological constant requires
\begin{equation}
\left|1-f(\rho)\right|^2 + \frac{\rho^2}{3} = 
\frac {(\Lambda_0 /k)^6}{a^2}\left(\frac{M_\mu}{M}\right)^2 \big(\mu /k)^{-4} \;,
\end{equation}
where 
\begin{equation}
f(\rho) = 1 + \frac{n-3}{2(n-1)} \left(\sqrt{1-\frac{8 (n-1)}{3 (n-3)^2} \rho^2}-1 \right) \;.
\end{equation}
Minimization of the potential yields the following results for $\mu$ and the different $F$-terms:
\begin{eqnarray}
\mu \a\simeq\a \Big(\!-\!\frac 3n a f(\rho)\Big)^{\frac 1{n-3}} \Lambda \;,\\
|F_\phi| \a\simeq\a \frac {\Lambda_{0}^3}{M^2}  \;,\\[2mm]
|F_\mu| \a\simeq\a  \frac {M}{M_\mu}\frac{\left|1-f(\rho)\right|}{\sqrt{\left(1-f(\rho)\right)^2 
+ {\rho^2}/{3}}} \frac{ \Lambda_{0}^3}{M^2}\,k \;,\\[-2mm]
|F_X| \a\simeq\a \frac{\rho}{\sqrt{\left(1-f(\rho)\right)^2 
+ {\rho^2}/{3}}}\frac{\left|\Lambda_{0}\right|^3}{M} \;.
\end{eqnarray}
The value of the parameter $\rho$ can be arbitrarily chosen between the minimal value $0$ and the maximal 
value $\sqrt{3/8}\sqrt{(n-3)^2/(n-1)}$. Taking $\rho $ and $f(\rho)$ of order one, we get 
$F_\phi \sim  F_\mu/k \sim F_X/M$. This implies that the contribution to $m_0^2$ in eq.~\eqref{m02bis} 
from radion and brane-to-brane mediations have the same magnitude and can compete with the contribution 
from anomaly mediation if $\mu/M \sim g^2/(4 \pi)^2$.  
Note that the localized kinetic term helps in making the radion-mediated contribution dominant 
compared to the brane-to-brane-mediated one, as we need to take $M_\mu < M$ in order to make it positive. 
More precisely, in order to significantly suppress the latter with respect to the former, one has to go to the 
regime in which the localized kinetic term gets close to its critical value. This is the analogue of a large 
localized kinetic term in the flat case. However, it is important to emphasize the in the flat case the suppression 
comes from the coefficients of the radiatively induced operators, whereas in the strongly warped case it comes from 
the scaling of the $F$-terms.

A more detailed study is needed to determine whether or not the scenario outlined here can be embedded in a fully viable model.
In order to do so we would have to choose a specific mechanism to generate the bulk superpotential (for instance 
by adding Golberger--Wise hypermultiplets in the bulk) and then check that the new sources of soft terms that are generated
in the specific model do not spoil the solution to the supersymmetric flavor problem. What we have just proven is the absence of
obstructions to obtain positive masses from gravitational loops plus anomaly mediation while having soft terms that scale like
\begin{eqnarray}
\a\a m_0 \sim m_{1/2} \sim \Big(\frac {g^2}{16 \pi^2}\Big)\,m_{3/2} \;,\\
\a\a m_{\rm radion} \sim (\mu /k)^{-1}\, m_{3/2} \;.
\end{eqnarray}
Overall, this scenario seems more promising  than the one presented in flat case, as we do not have the extra suppression of the 
radion-mediated contribution.  

\section{Summary and conclusions}\setcounter{equation}{0}

Anomaly mediation is very attractive as it is very model-independent and predictive, and solves the supersymmetric 
flavor problem. However, it predicts tachyonic sleptons, and therefore other contributions to the slepton soft masses are 
needed. In the context of sequestered models where the supersymmetry breaking sector and the SM sector are spatially 
separated in an extra-dimension, gravity loops can potentially provide such additional contributions. The transmission of 
supersymmetry breaking through these gravity loops is finite and calculable, because of its non-local  nature. Moreover, 
and for the same reason, it does not suffer from the flavor problem of the usual 4D gravity mediation of supersymmetry breaking. 
Recently, the contribution of gravitational loops to brane to brane and radion-mediated soft masses was computed in the 
case of a flat extra-dimension \cite{Rattazzi:2003rj,Buchbinder:2003qu}.  The result was partly disappointing as the 
contributions to the scalars soft masses squared were found to be negative in most cases. Fortunately, 
in ref.~\cite{Rattazzi:2003rj} it was found that the presence 
of a sizable localized gravitational kinetic term at the hidden brane induces
a positive contribution to the soft masses squared. 

Inspired by this result, in this paper we have calculated the effective K\"ahler potential in warped sequestered models. We 
used a superfield technique to perform our calculation. More precisely, we wrote the linearized bulk supergravity Lagrangian 
in term of $N=1$ superfields. Couplings between brane and bulk fields are then easily written down  using the known $N=1$ 
supergravity couplings.  We then performed a one loop supergraph calculation to find the effective K\"ahler potential 
valid in the 
presence of arbitrary localized kinetic terms for the supergravity fields. 
This calculation has the same flavor as the calculation of 
the Coleman-Weinberg potential and the result is given by
\begin{equation}
\Delta\Omega_{\rm 1-loop}=\frac{1}{2}\int \frac{d^4p}{(2\pi)^4}\sum_n \frac{-4}{p^2}\ln (p^2+\bar{m}_n^2) \;,
\end{equation}
where $\bar m_n$ are the masses of the KK gravitons in RS1, with arbitrary kinetic terms on both branes.

In general, we find that in the absence of localized kinetic terms for the supergravity multiplet, the gravitational contributions to 
the soft masses squared are negative, thus worsening the tachyonic slepton problem of anomaly mediation. In the limit of large warping, 
they go to zero, consistently with the fact that in this limit the theory becomes conformal. However, we find that adding a localized 
kinetic term on the IR brane can make the radion contribution to the soft masses positive. 

We emphasize that our formula is of a general nature. In particular, its generalization to higher co-dimensions should be easy to 
explore. This would only require the knowledge of the KK spectrum 
of the $J=2$ modes 
as a function of the moduli of the higher dimensional space in question, which is a relatively easy task. Also, the embedding of 
these moduli into superfields should be known, which should again be easy, if the low energy description of the model is known. 
We leave the exploration of this generalization for future work.

We also presented a simple way of understanding the form of the matter-dependent terms in the effective K\"ahler potential 
(radion and brane-to-brane contributions) as a function of the matter-independent one. The relevant observation is that for the 
low lying modes, that dominate our calculation, a shift in the position of the brane is equivalent to adding a localized kinetic 
term on the brane. Therefore, the radion-to-brane and brane-to-brane operators can be obtain from the matter independent 
effective K\"ahler potential by taking derivatives of the latter with respect to the radion.

Finally, we presented the effective description of a model where the radius is stabilized, and supersymmetry is broken in 
such a way that 
parametrically, the radion-mediated contribution to the soft masses squared can compete with the 
anomaly-mediated and brane-to-brane 
contributions, in a region of parameter space where the effective field theory description is valid. 
In such a model, phenomenologically viable 
supersymmetry breaking soft masses could arise from flavor universal and purely gravitational quantum effects. 

\section*{Acknowledgements}

We thank J.-P.~Derendinger, A.~Falkowski, M.~Luty and F.~Zwirner for useful discussions. 
This research was supported in part by the European Commission through a Marie Curie 
fellowship and the RTN contract  MRTN-CT-2004-503369.

\appendix

\section{Component Lagrangian}\setcounter{equation}{0}
\label{ap:components}

In this appendix, we describe the component form of the superfield Lagrangian \eqref{eq:bulklag}. 
Apart from the terms involving the prepotential $P_{\mathcal{T}}$ for the radion, the calculation 
is very similar to the one done in ref.~\cite{Linch:2002wg}, and we therefore refer the reader to 
that paper for more details. The first step is to chose a convenient Wess--Zumino gauge as in 
ref.~\cite{Linch:2002wg}, in which the bosonic components of the superfields are given by the following 
expressions:
\begin{eqnarray}
V_n \a=\a - \theta \sigma^m \thetabar \tilde h_{m n} + \theta^2 \thetabar^2 d_n \;, \nn \\[1mm]
\Psi_\alpha \a=\a \theta^\beta u_{\beta \alpha} 
+ \frac{i}{2} \thetabar_{\alphadot} {v_\alpha}^{\alphadot} 
+ \thetabar^2 \theta^\beta \Big(w_{\beta \alpha} 
+ \frac{1}{4} \sigma^n_{\beta \dot{\beta}} \partial_n {v_\alpha}^{\dot{\beta}}\Big) 
+ \theta^2 \thetabar_{\alphadot} \Big({y_\alpha}^\alphadot 
+ \frac{i}{2} {\sigmabar}^{n\alphadot \beta} \partial_n u_{\beta \alpha} \Big) \;, \nn \\[1mm]
P_\Sigma \a=\a - \theta \sigma^m \thetabar \tau^\Sigma_m + \theta^2 \thetabar^2 D_\Sigma \;, \nn \\[1mm]
\Sigma \a=\a \theta^2 \Big(D_\Sigma - \frac{i}{2} \partial_m \tau^\Sigma_m \Big) \;, \nn \\[-1mm]
P_{\mathcal{T}} \a=\a \rho_{\mathcal{T}} + \theta^2 \bar{t} + \bar \theta^2 t
- \theta \sigma^m \thetabar \tau^{\mathcal{T}}_m +\theta^2 \thetabar^2 
\Big(D_{\mathcal{T}} - \frac{1}{4} \square \rho_{\mathcal{T}}\Big) \;, \nn \\
{\mathcal{T}} \a=\a t + \theta^2 \Big(D_{\mathcal{T}} - \frac{i}{2} \partial_m \tau^{\mathcal{T}}_m\Big) 
+ i \theta \sigma^m \thetabar \partial_m t + \frac 14 \theta^2 \thetabar^2 \square t \;. \nn
\end{eqnarray}
In components this give:
\bea
\label{eq:fulllag}
\mathcal{L} = e^{-2 \sigma} \a\!\bigg\{\!\a 
\frac{1}{2} (\partial^n \tilde{h}_{nm})^2 
+ \frac{1}{2} (\partial^n \tilde{h}_{m n})^2 
- \frac{1}{2}(\partial_p \tilde{h}_{mn})^2 
+ \frac{1}{3} \tilde{h} \partial_m \partial_n \tilde{h}_{m n} \nn \\
\a\!\!\a + \frac{1}{6} (\partial_m \tilde{h})^2 
- \frac{1}{6} (\epsilon^{mnpq} \partial_p \tilde{h}_{m n})^2 
+ \frac{4}{3} d_n^2 + \frac{2}{3}\epsilon^{mnpq} \partial_p \tilde{h}_{mn} d_q \nn \\
\a\!\!\a + \frac{1}{4}\Big(\sigma^n_{\beta \alphadot} {w^\beta}_\alpha 
+ \sigma_{\alpha \dot{\beta}} {{\bar{w}}^{\dot{\beta}}}_\alphadot 
+ \text{Re} \partial_n v_{\alpha \alphadot} 
- e^{-\sigma} \partial_y \tilde{h}_{n \alpha \alphadot}\Big)^2 \nn \\
\a\!\!\a + \frac{1}{2}\text{Im}v_n \square \text{Im} v^n 
+ 2 e^{-\sigma} \text{Im} v_n \partial_y d^n 
+ \frac{i}{2}\text{Im} v^{\alpha \alphadot} \partial_{\beta \alphadot}{w^\beta}_\alpha 
- \frac{i}{2} \text{Im} v^{\alphadot \alpha} \partial_{\alpha \dot{\beta}} 
{\bar{w}^{\dot{\beta}}}_\alphadot \nn \\
\a\!\!\a + \frac{1}{2}\left|{w^\alpha}_\alpha 
+ \partial_n v_n - 3e^{-\sigma} \sigma^\prime t\right|^2 
+ \text{Re}t\, \partial_m \partial_n \tilde{h}_{mn} + 2 \left|y_n\right|^2 \nn \\
\a\!\!\a - 2 \Big[ \text{Re} y_n +\frac{i}{4} \sigmabar^{n\alphadot \beta}
\left(\partial_{\alpha \alphadot}{u^\alpha}_\beta 
- \partial_{\beta \dot{\beta}}{\bar{u}^{\dot{\beta}}}_\alphadot \right)
- \frac 14 e^{-\sigma} \left(\partial_y \tau^\Sigma_m 
+ 3 \sigma^\prime \tau^{\mathcal{T}}_m\right)\Big]^2 \nn \\
\a\!\!\a - \frac{3}{2} e^{- \sigma} \sigma^\prime \rho_{\mathcal{T}} 
\Big[2 \partial_p \text{Im} y_p - e^{-\sigma}\Big(\partial_y D_\Sigma 
+ 3 \sigma^\prime \big(D_{\mathcal{T}} 
- \frac{1}{4} \square \rho_{\mathcal{T}}\Big)\big)\Big] \nn \\
\a\!\!\a - D_{\mathcal{T}} D_\Sigma 
- \frac{1}{4}\partial_m \tau^{\mathcal{T}}_m \partial_m \tau^\Sigma_m 
- \frac{1}{3} D_\Sigma^2 - \frac{1}{12} (\partial_m \tau^\Sigma_m)^2 \bigg\} \;.
\eea
There are two independent sectors: one containing $h,t,v,w$ and $d$, and one 
containing $D_\Sigma,D_{\mathcal{T}},\tau^\Sigma_m,\tau^{\mathcal{T}}_m,y_n,
u_{\alpha \beta}$. In the former, $h,t$ and $v$ are physical fields, whereas 
$w$ and $d$ are auxiliary and must be integrated out. In the latter, none of 
the fields propagates. After eliminating all the auxiliary 
and non-propagating fields, one finally finds the following Lagrangian:
\bea
\mathcal{L} = M_5^3e^{-2 \sigma} \a\!\Big\{\!\a
(\partial^n h_{nm})^2 + h \partial_n \partial_m h_{nm} 
+ \frac{1}{2} (\partial_m h)^2 -\frac{1}{2} (\partial_p h_{mn})^2 \nn \\[-1mm]
\a\!\!\a + \frac{1}{2} e^{-2 \sigma} \left[\left(\partial_y h\right)^2 
- \left(\partial_y h_{mn}\right)^2\right] 
-\frac{1}{2}\left(\partial_n h_{my} - \partial_m h_{ny}\right)^2 \nn \\[1mm]
\a\!\!\a + 2 e^{-\sigma}\partial_m h_{ny} \partial_y h^{mn} 
- 2 e^{- \sigma} \partial^n h_{n y} \partial_y h
+ h_{yy} \partial^m \partial^n h_{mn} - h_{yy} \square h \nn \\[2mm]
\a\!\!\a  + 6 e^{-2 \sigma} \sigma^{\prime 2} h_{yy}^2 - 6 e^{- \sigma} h_{yy} \partial_n h_{n y} 
+ 3 e^{-2 \sigma} \sigma^\prime h_{yy} \partial_y h \nn \\
\a\!\!\a -\frac{1}{4}e^{2 \sigma}\left(\partial_n B_m - \partial_m B_n\right)^2 
 - \frac{1}{2} \left(\partial_m B_y - \partial_y B_m\right)^2 \Big\} \;,
\eea
where we have defined:
\begin{eqnarray}
h_{m n} \a=\a \frac 12 (\tilde h_{mn} + \tilde h_{nm}) - \frac 13 \eta_{mn} \tilde h \;,\;\;
h_{m y} = \frac{1}{2} \text{Re}\, v_m \;,\;\;
h_{yy} = \text{Re}\,t \;; \\
B_m \a=\a e^{-\sigma} \sqrt{\frac{3}{2}} \text{Im}\,v_m \;,\;\;
B_y = \sqrt{\frac{3}{2}} \text{Im}\,t \;.
\end{eqnarray}
This Lagrangian matches the expantion of the full RS1 Lagrangian to quadratic order
in the fluctuations, provided these are parametrized in the following way:
\begin{equation}
ds^2 = e^{-2 \sigma} \left(\eta_{m n} + h_{m n} \right)dx^m dx^n 
+ 2 e^{-\sigma} h_{m y} dx^m dy + (1 + h_{55}) dy^2
\end{equation}
In fact, a better parametrization of the fluctuations, that makes the zero mode 
manifest, is defined by:
\begin{equation}
ds^2 = e^{-2 \sigma (1 + T)} \left(\eta_{m n} + h_{m n}\right) dx^m dx^n 
+ 2 e^{-\sigma}h_{m y} dx^m dy + (1+T)^2 dy^2
\end{equation}
It can be checked that $T = \bar{T}(x)$ and $h_{mn} = \bar{h}_{mn}(x)$ are zero modes. 
This fact can be seen in our quadratic Lagrangian by replacing $h_{mn}$ by 
$\bar{h}_{mn}(x) - \sigma \bar{h}_{yy}(x)$, $h_{yy}$ by $\bar{h}_{yy}(x)$, 
$B_y$ by $\bar{B}_y(x)$, setting the odd field $h_{m y}$ and $B_m$ to zero,
and integrating over $y$. We get in this way:
\begin{eqnarray}
{\cal L} \a=\a M_5^3 \bigg\{\frac{1 \!-\! e^{-2 k \pi R}}{k} 
\Big[\left(\partial^n \bar h_{n m}\right)^2\! + \bar h \partial^m \partial^n \bar h_{mn} 
+ \frac{1}{2} \left(\partial_m \bar h \right)^2\! - \frac{1}{2}
\left(\partial_p \bar h_{mn}\right)^2\! 
-\frac{1}{2} \left(\partial_m \bar B_y \right)^2 \Big] \nn \\
&& \hspace{20pt} +\, \pi R\, e^{-2 k \pi R }\Big[ 
\left( \partial^n \partial^m \bar h_{mn}-\square \bar h\right) \bar h_{yy} 
+ 3 k \pi R\, \bar h_{yy} \square \bar h_{yy}\Big] \bigg\} \;.
\end{eqnarray}


\end{document}